\documentclass[twocolumn,trackchanges]{aastex62}

\shorttitle{PPD Turbulence Sample}
\shortauthors{Flaherty et al.}

\begin{document}

\title{Measuring turbulent motion in planet-forming disks with ALMA: A detection around DM Tau and non-detections around MWC 480 and V4046 Sgr}



\correspondingauthor{Kevin Flaherty}
\email{kmf4@williams.edu}

\author[0000-0003-2657-1314]{Kevin Flaherty}
\affil{Department of Astronomy and Department of Physics\\
Williams College, Williamstown, MA 01267 USA}

\author{A. Meredith Hughes}
\affil{Van Vleck Observatory, Astronomy Department, Wesleyan University \\
96 Foss Hill Drive \\
Middletown, CT 06459, USA}

\author{Jacob B. Simon}
\affil{Department of Physics and Astronomy, Iowa State University, Ames, IA, 50010, USA}
\affil{Department of Space Studies, Southwest Research Institute\\
Boulder, CO 80302}
\affil{JILA, University of Colorado and NIST\\
440 UCB, Boulder, CO 80309}

\author{Chunhua Qi}
\affil{Harvard-Smithsonian Center for Astrophysics\\
60 Garden St\\
Cambridge, MA 02138}

\author{Xue-Ning Bai}
\affil{Institute for Advanced Study\\
Tsinghua University\\
Beijing 100084, China}
\affil{Tsinghua Center for Astrophysics\\
Tsinghua University\\
Beijing 100084, China}

\author{Alyssa Bulatek}
\affil{Department of Physics \& Astronomy \\
Macalester College\\
1600 Grand Avenue, Saint Paul, MN 55105}

\author{Sean M. Andrews}
\affil{Harvard-Smithsonian Center for Astrophysics\\
60 Garden St\\
Cambridge, MA 02138}

\author{David J. Wilner}
\affil{Harvard-Smithsonian Center for Astrophysics\\
60 Garden St\\
Cambridge, MA 02138}

\author[0000-0001-7157-6275]{\'{A}gnes K\'{o}sp\'{a}l}
\affil{Konkoly Observatory, Research Centre for Astronomy and Earth Sciences, \\
Konkoly-Thege Mokl\'{o}s \'{u}t 15-17, 1121 Budapest, Hungary}
\affil{Max Planck Institute for Astronomy, K\"{o}nigstuhl 17, D-69117 Heidelberg, Germany}
\affil{ELTE E\"otv\"os Lor\'and University, Institute of Physics, P\'azm\'any P\'eter s\'et\'any 1/A, 1117 Budapest, Hungary}

\begin{abstract}
Turbulence is a crucial factor in many models of planet formation, but it has only been directly constrained among a small number of planet forming disks. Building on the upper limits on turbulence placed in disks around HD 163296 and TW Hya, we present ALMA CO J=2-1 line observations at $\sim0\farcs3$ (20-50 au) resolution and 80 ms$^{-1}$ channel spacing of the disks around DM Tau, MWC 480, and V4046 Sgr. Using parametric models of disk structure, we robustly detect non-thermal gas motions around DM Tau of between 0.25 c$_s$  and 0.33 c$_s$, with the range dominated by systematic effects, making this one of the only systems with directly measured non-zero turbulence. Using the same methodology, we place stringent upper limits on the non-thermal gas motion around MWC 480 ($<$0.08c$_s$) and V4046 Sgr ($<$0.12 c$_s$). The preponderance of upper limits in this small sample, and the modest turbulence levels consistent with dust studies, suggest that weak turbulence ($\alpha\lesssim10^{-3}$) may be a common, albeit not universal, feature of planet-forming disks. We explore the particular physical conditions around DM Tau that could lead this system to be more turbulent than the others. 

\end{abstract}



\section{Introduction} \label{sec:intro}
Understanding the connection between disks around young stars and the planets formed within these environments requires knowledge about a wide range of physical processes, as well as how these processes vary between different stellar systems. One of the key disk properties is the non-thermal, non-Keplerian turbulent motion, which theoretically influences a range of processes ranging from the growth of grains from micron-sizes up to planets \citep[e.g.][]{orm07,bir10} and the orbital evolution of fully-formed planets \citep[e.g.][]{nel04,paa11,fun14} to the chemical \citep{sem11,fur14,xu17} and angular momentum \citep[e.g.][]{lyn74} evolution of the disk. 

Efforts to broadly characterize the turbulence among a large sample of young stellar objects have relied on indirect techniques. Assuming that a planet-forming disk evolves through viscous processes, in which angular momentum is transported outwards while mass is transported inwards, leads to a predicted connection between accretion rate, disk mass, disk radius, and time \citep{har98} that can be used to constrain turbulence. In the context of an $\alpha$-disk model of viscosity \citep{sha73}, in which the viscosity ($\nu$) is equal to the product of $\alpha$, the local sound speed ($c_s$), and the gas pressure scale height ($H$), a number of studies have examined various predictions of viscous evolution using samples of $\sim$30-50 sources \citep{ste98,mul17,lod17,raf17,ans18,naj18} and have found a wide range of $\alpha$ values (10$^{-4}$ - 0.1). \citet{mul12} rely on the connection between turbulence and the settling of dust grains \citep[e.g.][]{you07}, and use the average spectral energy distributions of Herbig stars, T-Tauri stars, and brown dwarfs to find that $\alpha\sim10^{-4}$ best reproduces the average emission profile. High resolution continuum imaging of dust structures also provides constraints on non-thermal motion through the sharpness and morphology of both dark gaps and bright rings \citep[e.g.][]{pin16,hua18,dul18}.

Gas motion provides a more direct measure of the turbulence, but observational constraints are limited to small samples. \citet{hug11} studied the disks around HD 163296 and TW Hya, while \citet{gui12} studied the disk around DM Tau. In the ALMA era, TW Hya and HD 163296 have been revisited \citep{fla15,tea16,fla17,fla18,tea18a}, finding upper limits on the turbulence of $\lesssim$5-10\%\ of the local sound speed for gas at distances of $\sim$30 au and larger from the central star. Near-infrared observations, more sensitive to gas within a few au of the central star, have found evidence of non-thermal broadening components comparable to the local sound speed among samples of similar sizes \citep{car04,naj09,dop11,ile14}. 

Here we build on our previous efforts to measure turbulence directly from ALMA observations of molecular gas emission lines to include the disks around DM Tau, MWC 480, and V4046 Sgr. While still representing a modest sample, these sources were chosen to cover a range of accretion rates, X-ray luminosities, and far-ultraviolet (FUV) luminosities. This parameter space is particularly relevant for the magneto-rotational instability \citep{bal98}. The MRI relies on the coupling between gas and magnetic field to destabilize a rotating disk to generate the correlated turbulent fluctuations that can transport angular momentum and drive accretion through the disk. The outcome of the MRI, especially the level of turbulence, is sensitive to the level of disk ionization \citep[e.g.][]{sim18}. Therefore, observationally establishing a relation between turbulence and other disk parameters offers important physical insight into the microphysical processes governing disk evolution.

In section 2 we present the ALMA observations of these three systems, and in section~\ref{models} we discuss our framework for modeling the CO emission. We present our finding of weak turbulence around MWC 480 and V4046 Sgr, and non-zero turbulence around DM Tau in section~\ref{sec_results}, while in section~\ref{discussion} we examine the evidence that weak turbulence may be common among planet-forming disks, as well as the physical conditions that influence the difference in the measured level of turbulence between DM Tau and the other systems. 

\section{Observations}
ALMA band 6 observations of the disks around DM Tau, MWC 480, and V4046 Sgr were taken as part of project 2016.1.00724.S. These observations were split into short and long baseline components in order to capture the full spatial extent of the CO emission (out to $\sim$6$\arcsec$) at high spatial resolution ($\sim$0$\farcs$3). Observations were taken between December 2016 and July 2017, with baselines ranging from 15 m to 2.6 km (Table~\ref{observing_table}). 

The spectral windows were split between a continuum window centered at 232 GHz, and three windows centered on CO(2-1), $^{13}$CO(2-1), and C$^{18}$O(2-1) respectively. This paper focuses on CO(2-1), while future work will consider the isotopologues. For the spectral lines, the correlator was configured for 960 channels, each 61.035 kHz wide ($\sim$80 m s$^{-1}$).  The standard ALMA reduction scripts were used to produce calibrated visibilities. For each set of observations, two rounds of phase self-calibration and one round of amplitude self-calibration was performed, based on the high S/N continuum data with the exception of the short-baseline DM Tau observations where the amplitude self-calibration was excluded. The spectra were binned down by a factor of 2 (0.16 km s$^{-1}$).

Uncertainties in the flux calibration will result in small variations in the flux between the two sets of observations. To remove this effect, we compare the continuum visibilities with baselines $<$200 k$\lambda$ among the two observational epochs. We find that the long baseline observations are 1.06, 1.04, and 0.93 times brighter than the short baselines observations for DM Tau, MWC 480, and V4046 Sgr respectively; the size of the amplitude differences is consistent with that expected from the uncertainty in the ALMA calibration \citep{but12}. These scale factors are applied to the model images before fitting to the long baseline data.

The positional offset of the disk between the two epochs can vary based on e.g. errors in the proper motion correction or errors in the astrometric accuracy of ALMA. We estimate the positional offset between the short and long baselines observations by fitting an elliptical Gaussian to the continuum observations, and apply a phase offset to each epoch to correct the disk center to the phase center. For MWC 480 the positional offsets are 73 mas, -86 mas in RA and Dec for the short baseline data, and 56 mas, -3 mas for the long baseline data. For the DM Tau disk, the offsets in RA and Dec are 230 mas, 3.6 mas for the short baseline data and 230 mas, -2.2 mas for the long baseline data. No significant offset between the short and long-baseline data was found for V4046 Sgr, although for both data sets the center of the disk was offset by 270 mas in RA and 17 mas in Dec from the phase center, which is included in the models. The observed locations of the disks are consistent with the stellar positions now available in the GAIA DR2 database \citep{gaia18}. The amplitude differences between the short and long baselines are unaffected by these positional offsets.

Cleaned images were generated by combining the short and long baseline visibilities. Using natural weighting results in $\sim$0$\farcs$3-0$\farcs$5 beam with an rms of 2-4 mJy/beam per 0.16 km s$^{-1}$ wide channel.

\begin{deluxetable*}{ccccccc}[h]
\tabletypesize{\scriptsize}
\tablewidth{0pt}
\tablecaption{Observing Summary\label{observing_table}}
\tablehead{\colhead{Star} & \colhead{Date} & \colhead{Time on source} & \colhead{Baselines} & \colhead{PWV} & \colhead{Calibrators} & \colhead{Phase Center}}
\startdata
DM Tau & 2016-12-27 & 10 min & 15-459 m & 1.5 mm & J0510+1800 (bandpass, phase), J0423-0120 (flux) & 04:33:48.732781 +18:10:09.67744 \\
DM Tau & 2017-07-05 & 32 min & 21-2647 m & 0.61 mm & J0510+1800 (bandpass, phase), J0423-0120 (flux) & 04:33:48.733234 +18:10:09.66785\\
MWC 480 & 2016-12-27 & 12 min & 15-459 m & 1.65 mm & J0510+1800 (bandpass,flux), J0438+3004 (phase) & 04:58:46.271217 +29:50:36.52973\\
MWC 480 & 2016-10-27 & 39 min & 18-1124 m & 0.77 mm & J0510+1800 (bandpass,flux), J0438+3004 (phase) & 04:58:46.271713 +29:50:36.53469\\
V4046 Sgr & 2017-04-08 & 8 min & 15-390 m & 1.18 mm & Titan (flux), J1924-2914 (bandpass), J1826-2924 (phase) & 18:14:10.468875 -32:47:35.43708 \\
V4046 Sgr & 2017-05-03 & 27 min & 15-1110 m & 0.38 mm & J1924-2914 (flux,bandpass), J1826-2924 (phase) & 18:14:10.468886 -32:47:35.44081\\
\enddata
\end{deluxetable*}

\section{Models\label{models}}
The modeling code used to derive the turbulence has been described previously in \citet{fla15}, \citet{fla17}, and \citet{fla18}, and is based on earlier work by \citet{ros13} and \citet{dar03}. We summarize the relevant equations below, including the key free parameters and the assumptions made for each source. 

The surface density is assumed to follow a power law, with an exponential tail, as expected for a viscously evolving disk \citep{lyn74,har98}.
\begin{equation}
\Sigma_{\rm gas}(r) = \frac{M_{\rm gas}(2-\gamma)}{2\pi R^2_c}\left(\frac{r}{R_c}\right)^{-\gamma}\exp\left[-\left(\frac{r}{R_c}\right)^{2-\gamma}\right],
\end{equation}
where $M_{\rm gas}$, $R_c$ and $\gamma$ are the gas mass (in M$_{\odot}$), critical radius (in au) and power law index respectively. The disk extends from $R_{\rm in}$ to 1000 au. 

The temperature structure is defined as a power law with radius, with a vertical gradient connecting the cold midplane with the warm atmosphere. 
\begin{eqnarray}
T_{\rm mid} & = & T_{\rm mid0}\left(\frac{r}{150\ \rm au}\right)^{q}\\
T_{\rm atm} & = & T_{\rm atm0}\left(\frac{r}{150\ \rm au}\right)^{q}\\
T_{\rm gas}(r,z) & = & \left\{
\begin{array}{ll}
T_{\rm atm} + (T_{\rm mid}-T_{\rm atm})(\cos\frac{\pi z}{2Z_q})^{2} & \mbox{if $z < Z_q$} \\
T_{\rm atm} & \mbox{if $z \ge Z_q$}
\end{array}
\right.\\
Z_q & = & Z_{q0} (r/150\ {\rm au})^{1.3}
\end{eqnarray}
The parameter $Z_q$ is the height above the midplane at which the gas temperature reaches its maximum value. Once the temperature structure and surface density profile have been specified, the hydrostatic equilibrium calculation is performed at each radius to derive the gas volume density at each height above the midplane. The velocity field is assumed to be Keplerian, with corrections for the height above the midplane and the pressure support of the gas, as in \citet{ros13}. 

The line profile is a Gaussian whose width is set by the thermal and non-thermal motions. We associate the non-thermal term with turbulence ($\delta v_{\rm turb}$) although we consider other sources of non-Keplerian, non-thermal motion in section~\ref{sec:zonal}. We focus on cases in which the non-thermal term is proportional to the local isothermal sound speed:
\begin{equation}
\Delta V = \sqrt{\left(2k_BT(r,z)/m_{CO}\right)(1+\delta v_{\rm turb}^2)}.
\end{equation}
This parameterization is physically motivated by numerical simulations that predict that the turbulence scales with the local sound speed \citep[e.g.][]{shi14,sim15,flo17}.

One important component in fitting the CO observations is the midplane temperature. As demonstrated in \citet{fla18}, an overestimate of the midplane temperature can lead to an underestimate of the turbulence. The data themselves do not provide strong constraints on the midplane temperature; C$^{18}$O(2-1) has a strong degeneracy between $T_{\rm mid}$ and the CO abundance, while CO(2-1), unlike other systems \citep[e.g.][]{ros13,pin18a,dul19}, does not have enough sight lines at low optical depth that reach to the midplane. 

For DM Tau, the value of $T_{\rm mid0}$ is chosen such that CO freeze-out at the midplane ($T_{\rm mid}$=19 K) occurs at 70 au, as predicted by the chemical models of \citet{loo15}. To preserve the location of CO freeze-out while the temperature structure, through the parameter $q$, is varied during the model fitting, the exact value of the $T_{\rm mid0}$ parameter is defined based on the value of $q$ (i.e. $T_{\rm mid0}$=19(70/150)$^{-q}$). For MWC 480, we use the temperature structure derived by \citet{pie07} based on observations of $^{13}$CO, which corresponds to $T_{\rm mid0}$=18.8 K. For the disk around V4046 Sgr, we use a midplane normalization based on a D'Alessio model fitting of the SED, which results in $T_{\rm mid0}$=12 K (C. Qi. private communication). In addition to these fiducial values, we also consider lower $T_{\rm mid0}$ values to examine the effect of a different assumed value of the midplane temperature on the turbulence estimates.

In our initial attempts at fitting the CO emission, we found that the models were unable to simultaneously reproduce the emission at radii $\lesssim$200 au and $\gtrsim$200 au. This likely reflects complexities in the CO freeze-out/desorption process that are not included in our initial model prescription; we implicitly assume that thermal desorption and photo-dissociation set the CO molecular layer, while recent chemical models have shown that non-thermal desorption effects are important around IM Lup \citep{obe15} and TW Hya \citep{tea17}. Similarly, extended cold CO emission has been observed around AS 209 \citep{hua16}, and TW Hya \citep{sch16}. To account for non-thermal desorption of CO, we re-introduce CO to the disk in the region between $N_{H_2}=1.3$ and 4.8 $\times$10$^{21}$ cm$^{-2}$ when $T_{\rm gas}<$20 K, as described in Appendix~\ref{CO_photo_desorption}. 

In our fiducial models we allow $T_{\rm atm0}$, $q$, $R_c$, $i$, $\delta v_{\rm turb}$, and $R_{\rm in}$ to vary, along with the systemic velocity of the disk ($v_{\rm sys}$). Table~\ref{fixed_params} lists the fixed parameters for each source. All distances are based on GAIA DR2 parallax measurements \citep{gaia18,lin18}. The position angle of the disk is determined based on initial fitting of the system (PA=154.8$^{\circ}$, 147.8$^{\circ}$, 75.65$^{\circ}$ for DM Tau, MWC 480, and V4046 Sgr respectively), and was kept fixed due to its independence from the other model parameters, while $\gamma$ is always fixed at 1. For the disk around V4046 Sgr, we use the sum of the masses of the two stars to set the Keplerian motion of the disk. The sign on the inclination is adjusted so as to minimize asymmetric residuals (Appendix~\ref{inclination_sign}). Using GAIA DR2 distances \citet{sim19} derive stellar masses for MWC 480 and DM Tau of 2.11$\pm$0.01 M$_{\odot}$ and 0.55$\pm$0.02 M$_{\odot}$ respectively. Utilizing these updated masses will most strongly affect the derived inclination \citep[e.g.][]{cze16} while their affect on turbulence is likely to be small, especially for DM Tau, and will be investigated in more detail in future work.

The posterior distributions for each parameter are estimated using the MCMC routine EMCEE \citep{for13}, based on the Affine-Invariant algorithm originally proposed in \citet{goo10}. The uncertainties in the data are estimated based on the calculated dispersion among baselines of similar distances in line-free channels. Typical MCMC chains consist of 80 walkers and 800 steps, with convergence on the final solution occurring within 300 steps. The first 500 steps are removed as burn-in, after which the median of the posterior distributions varies by $<$1\%. We use linear spacing in all parameters except $R_c$, where we fit $\log(R_c)$. We assume uniform priors that simply restrict parameters to physically realistic values (e.g. $\delta v_{\rm turb}>$0, $T_{\rm atm0}>$0, $R_{\rm in}>0$).

 \begin{deluxetable*}{cccccccc}
\tabletypesize{\scriptsize}
\tablecaption{Fixed Model Parameters\label{fixed_params}}
\tablehead{\colhead{Star} & \colhead{Stellar Mass (M$_{\odot}$)} & \colhead{M$_{\rm gas}$ (M$_{\odot}$)} & \colhead{X$_{\rm CO}$} & \colhead{Distance (pc)} & \colhead{PA} & \colhead{$Z_{q0}$ (au)} & \colhead{$\gamma$}}
\startdata
DM Tau & 0.54 (1) & 0.04 (2,3) & 2$\times$10$^{-5}$ (3) & 144.5 $\pm$ 1.1 (4) & 154.8$^{\circ}$ & 70 & 1\\
MWC 480 & 1.85 (5) & 0.046 (6) & 10$^{-4}$ & 161.1 $\pm$ 2.0 (4) & 147.8$^{\circ}$ & 50 & 1\\
V4046 Sgr & 1.75 (7) & 0.09 (8) & 3$\times$10$^{-6}$ (7) & 72.26 $\pm$ 0.34 (4) & 75.7$^{\circ}$ & 35 & 1\\
\enddata
\tablenotetext{}{References: (1) \citet{sim00}, (2) \citet{and11}, (3) \citet{mcc16}, (4) \citet{lin18,bai18}, (5) \citet{pie07}, (6) This work (7) \citet{ros12},  (8) \citet{ros13b}}
\end{deluxetable*}

\section{Results\label{sec_results}}
\subsection{Non-zero turbulence around DM Tau}
\begin{figure*}
\includegraphics[scale=.4]{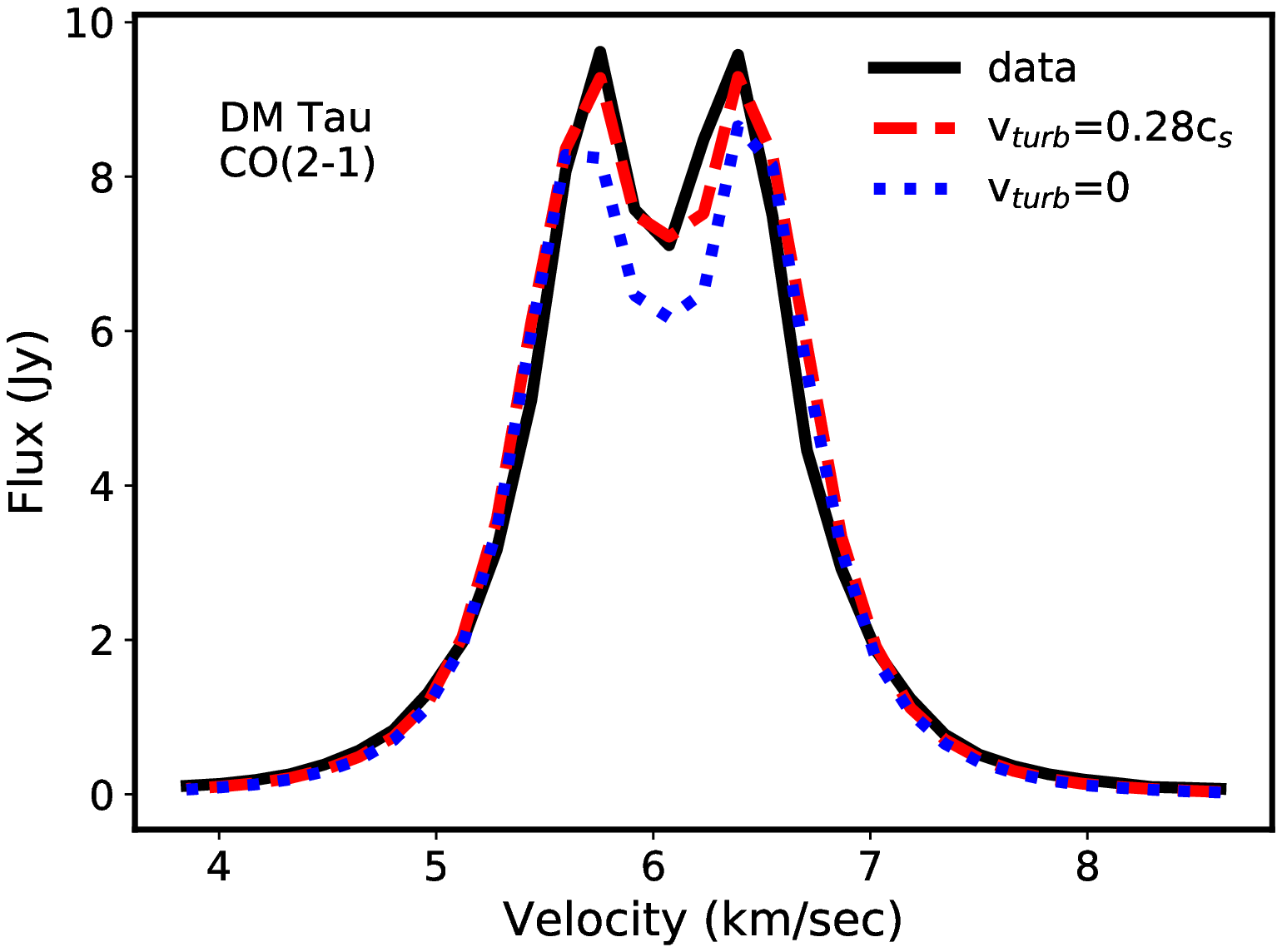}
\includegraphics[scale=.28]{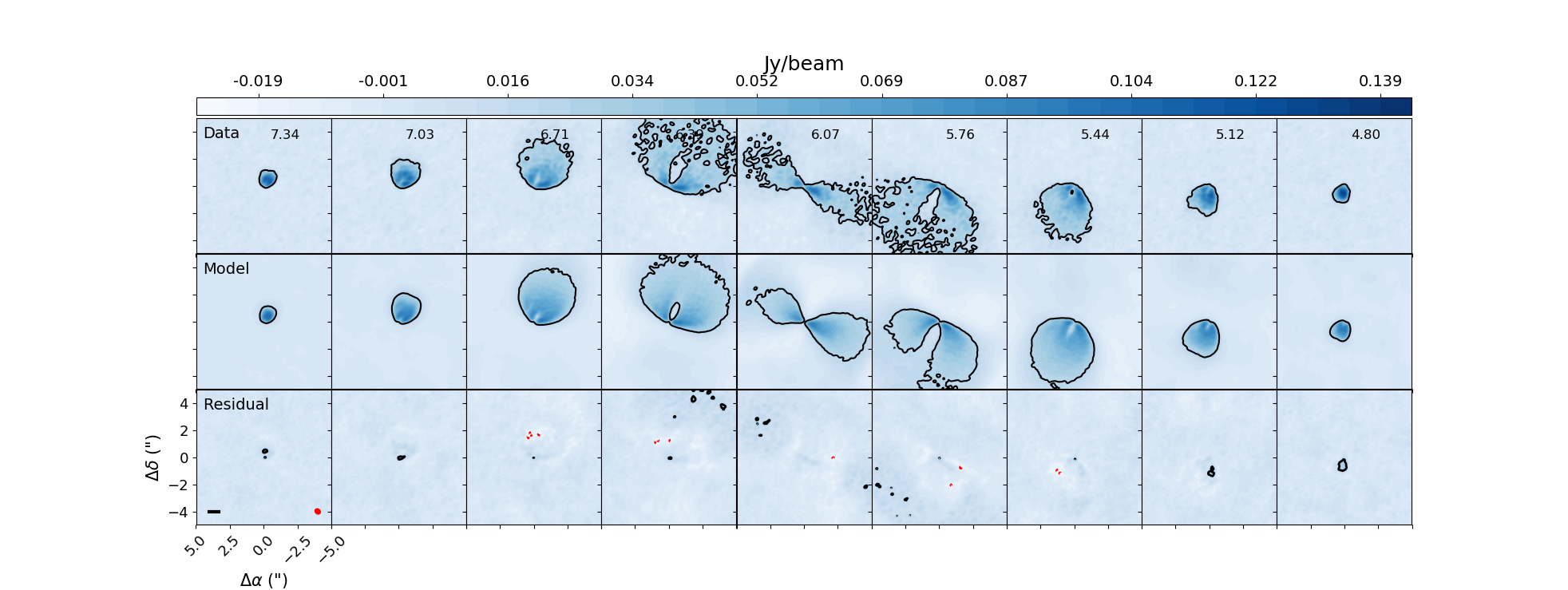}
\caption{(Left:) Spectra of the DM Tau CO(2-1) emission (black line) and the model defined by the median of the PDFs (red-dashed line). The blue dotted line indicates a model fit in which turbulence has been fixed at zero, while the rest of the parameters are allowed to vary. The zero-turbulence fit is significantly worse than the non-zero turbulence fit.
(Right:) Channel maps of the data (top row), model with non-zero turbulence (middle row), and the imaged residuals (bottom row). For the data and model frames the black contours mark the 5$\sigma$ (=22mJy/beam$\sim$17\%\ of the peak flux) boundary. In the residual frames, positive (black) contours and negative (red dashed) contours are included at multiples of 5$\sigma$. The bottom left panel includes a 100 au scale bar and the synthesized beam. Here we only show the model with non-zero turbulence, which is able to reproduce nearly all of the CO emission from around DM Tau. \label{dmtau_results}}
\end{figure*}

Unlike in the disks around HD 163296 and TW Hya, we find that a non-zero turbulence provides the best fit to the disk around DM Tau, with a 99.7\% confidence interval on the turbulent motion of 0.279$^{+0.005}_{-0.004}$c$_s$ (Table~\ref{results}), corresponding to velocities of 60-80 m s$^{-1}$. Model spectra, and residuals generated from the difference between model and data visibilities, are shown in Figure~\ref{dmtau_results}. Full posterior distributions, along with the progression of the walkers through the chains, are shown in appendix~\ref{pdfs}. The non-zero turbulence model spectrum is well matched to the data, and outside of a persistent unresolved (diameter $<$ 40 au) positive residual at the center, indicating that the model underestimates the data, the channel maps do not show strong residuals. As discussed below, similar central residuals are seen around MWC 480 and V4046 Sgr and may be due to deviations from our simple prescriptions for density and temperature, or a sign of changing CO abundance with radius.

The MCMC modeling indicates that a non-zero turbulence exists at a highly significant level. To verify this conclusion, we perform an additional MCMC run with the turbulence fixed at zero. This results in a model that systematically underestimates the flux of the disk (Figure~\ref{dmtau_results}). Comparing the median models from both trials using the Akaike Information Criterion \citep{aka74} the difference is confirmed, with the zero turbulence model being worse than the non-zero turbulence model at a level that is much greater than 10$\sigma$. We also can confirm the non-zero turbulence by fitting to the un-binned spectra while allowing turbulence to vary. Again we find non-zero turbulence (0.252$^{+0.003}_{-0.004}$c$_s$) confirming our fit to the binned spectrum.

We find an uncertainty on the turbulence at the $\sim$2\%\ level, but this does not include the systematic uncertainty on the amplitude calibration, which may affect the derived temperature structure, and in turn the turbulent broadening. The 4-7\%\ variation in amplitude measured between the short and long baseline observations of the three systems under study here is consistent with some uncertainty in the flux calibration of these data. To test the robustness of our result in the face of the amplitude calibration uncertainty, we run additional trials in which the model is systematically scaled by $\pm$20\%, simulating a true disk flux that is 20\%\ fainter/brighter than is actually observed. Again we find significant turbulence, with $\delta v_{\rm turb}$=0.300$\pm$0.005c$_s$ for a faint disk and $\delta v_{\rm turb}$=0.257$^{+0.004}_{-0.005}$c$_s$ for a bright disk. These results differ in the gas temperature with $T_{\rm atm0}$=21.26$^{+0.09}_{-0.08}$ K for the faint disk and $T_{\rm atm0}$=28.08$^{+0.15}_{-0.12}$ K for the bright disk, which in turn affects the exact value of the sound speed. This anti-correlation between T$_{\rm atm0}$ and $\delta v_{\rm turb}$ is also seen in the fiducial model posterior distribution (Figure~\ref{dmtau_pdfs}). The range in $\delta v_{\rm turb}$ values between the systematic uncertainty models is larger than in the fiducial model since the amplitude calibration uncertainty allows for a larger range in $T_{\rm atm0}$. The anti-correlation between $\delta v_{\rm turb}$ and $T_{\rm atm0}$ does indicate that uncertainty in the gas temperature structure influence our turbulence constraints.

As discussed earlier, the assumed midplane temperature also influences the final turbulence result. We run an additional MCMC ensemble in which $T_{\rm mid0}$ has been reduced to 11 K, as suggested by observations of N$_2$H$^+$ \citep{qi19}, and lower than employed in the chemical models of \citet{loo15}. Again we find non-zero turbulence, at a level of $\delta v_{\rm turb}$=0.328$\pm$0.004c$_s$. The anti-correlation between midplane temperature and turbulence arises because both influence the strength of the emission at the edges of the emitting regions within a given channel map \citep{fla18}. These regions have low optical depth and probe the gas temperature closer to the midplane; increasing the midplane temperature will increase the intensity of the emission from these edges. At the same time, increasing turbulence will increase the amount of emission that ``bleeds over" from nearby channels, leading to an increase in the optical depth and emission that arises from higher and warmer layers. As a result, a given intensity along the edges of the emission within a specific channel can be achieved with either a cold midplane and strong turbulence, or weaker turbulence and a warmer midplane. When combined with the results from the amplitude calibration trials, the low $T_{\rm mid0}$ trial indicates that systematic uncertainties dominate over statistical uncertainties in our measurement of turbulence. Based on the sample of trials run here, we find $\delta v_{\rm turb}=0.25-0.33$c$_s$, significantly larger than zero, but with a larger uncertainty than indicated by any individual trial. 

\begin{figure*}
\center
\includegraphics[scale=.5]{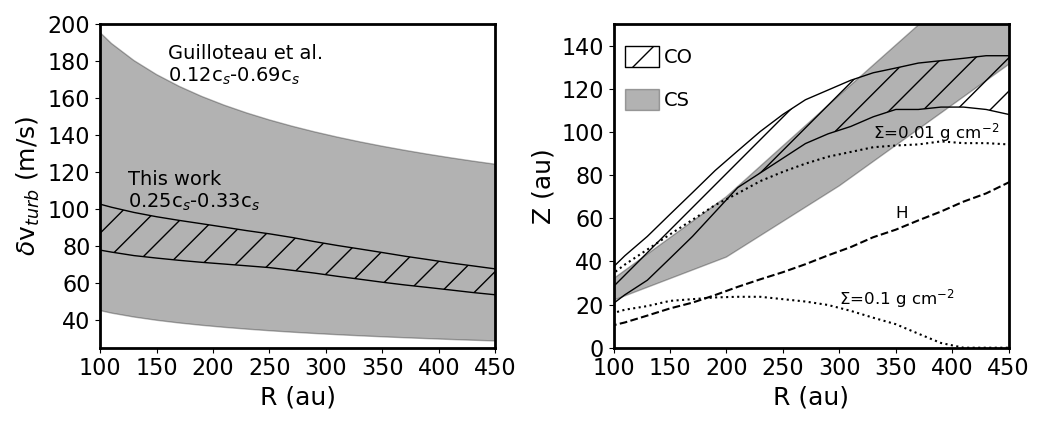}
\caption{(Left) Turbulence levels as a function of radius in the disk around DM Tau, as derived from CS (\citet{gui12}, wide grey band) and CO (this work, narrow hatched band). For the \citet{gui12} results we show the 3$\sigma$ range, including the uncertainty on both the derived turbulence and the gas temperature (when deriving turbulence as a function of the sound speed), while in the results from this work we include systematic uncertainties. Our results are consistent with the previous measurements, albeit with smaller uncertainties. (Right) Molecular emitting regions for CO (this work) and CS \citep{gui12}. Both molecules are probing similar regions, at one to three pressure scale heights (H) above the midplane. \label{turb_struct_dmtau}}
\end{figure*}

Previous analysis of CS(3-2) and CS(5-4) \citep{gui12} found significant turbulence around DM Tau, at a level of 120 m s$^{-1}$, with CS being used instead of CO because of its larger molecular weight and hence smaller thermal broadening. When including the uncertainty on the turbulence velocity and the temperature structure, this translates to a 3$\sigma$ range of (0.12-0.69)c$_s$, consistent with our measured range of (0.25-0.33)c$_s$ when accounting for systematic uncertainties (Figure~\ref{turb_struct_dmtau}). The higher signal-to-noise and higher spatial resolution ALMA data provide a tighter constraint on the turbulence around DM Tau.

Interpreting these results in the context of theoretical models requires knowledge of the emitting region of the molecular emission. Figure~\ref{turb_struct_dmtau} shows the emitting region of CO derived from our models as well as that of CS from \citet{gui12}; the emission from both molecules arises from between two and three pressure scale heights above the midplane. The emitting regions are at low column densities, where far ultraviolet photons are expected to ionize the disk, making it more susceptible to MRI \citep{per11}. At $\delta v_{\rm turb}=0.25-0.33$c$_s$, the turbulence is at the low end for model predictions from MRI \citep{sim15,sim18}, consistent with gravito-turbulence \citep{for12,shi14,shi16}, and at the high end for the vertical shear instability \citep{ric16,flo17}. Distinguishing between these different models for the source of the turbulence may require a measure of the vertical structure of the turbulence \citep{sim15}, more detailed knowledge of the physical conditions within the outer disk \citep{lyr19,sim18}, or complementary observations of other disk properties that are influenced by the turbulence \citep[e.g. dust settling][]{flo17}

\begin{deluxetable*}{ccccccccc}[h]
\tabletypesize{\scriptsize}
\tablewidth{0pt}
\tablecaption{CO(2-1) Model results\label{results}}
\tablehead{\colhead{Model} & \colhead{q} & \colhead{$\log (R_c (\rm au))$} & \colhead{$\delta v_{\rm turb}$ (c$_s$)} & \colhead{$T_{\rm atm0}$ (K)} & \colhead{$T_{\rm mid0}$ (K)} & \colhead{inclination ($^{\circ}$)}  & \colhead{$R_{\rm in}$ (au)} & \colhead{$v_{\rm sys}$ (km s$^{-1}$)}}
\startdata
\cutinhead{DM Tau}
Fiducal & -0.371$^{+0.003}_{-0.002}$ & 2.444$_{-0.003}^{+0.004}$ & 0.279$^{+0.005}_{-0.004}$  & 24.68$^{+0.11}_{-0.09}$ & 14.3\tablenotemark{a} & 36.0$^{+0.12}_{-0.09}$ & $<$9 & 6.017$\pm0.001$\\ 
sys=1.2 & -0.344$\pm0.003$ & 2.452$^{+0.003}_{-0.004}$ & 0.257$^{+0.004}_{-0.005}$  & 28.08$^{+0.15}_{-0.12}$ & 14.6\tablenotemark{a} & 36.21$^{+0.12}_{-0.13}$ & $<$9 & 6.018$^{+0.001}_{-0.002}$\\ 
sys=0.8 & -0.403$^{+0.003}_{-0.002}$ & 2.44$\pm0.01$ & 0.300$\pm0.005$  & 21.26$^{+0.09}_{-0.08}$ & 14.0\tablenotemark{a} & 35.81$^{+0.12}_{-0.13}$ & $<$9 & 6.016$\pm0.001$\\ 
highres & -0.352$^{+0.001}_{-0.002}$ & 2.440$\pm0.002$ & 0.252$^{+0.003}_{-0.004}$  & 24.92$^{+0.08}_{-0.07}$ & 14.5\tablenotemark{a} & 35.99$^{+0.09}_{-0.08}$ & $<$9 & 6.045$\pm0.001$\\ 
$\delta v_{\rm turb}$=0 & -0.352$^{+0.004}_{-0.002}$ & 2.456$^{+0.004}_{-0.003}$ & 0\tablenotemark{b} & 27.54$\pm0.09$ & 14.5\tablenotemark{a} & 37.73$^{+0.16}_{-0.11}$ & $<$9 & 6.013$^{+0.002}_{-0.001}$\\ 
low $T_{\rm mid0}$ & -0.310$^{+0.002}_{-0.003}$ & 2.469$^{+0.004}_{-0.005}$ & 0.328$\pm$0.004 & 25.57$^{+0.09}_{-0.07}$ & 11.0\tablenotemark{c} & 35.86$\pm0.11$ & $<$9 & 6.019$\pm0.001$\\ 
\cutinhead{MWC 480}
Fiducial & -0.467$\pm0.002$ & 2.161$^{+0.004}_{-0.001}$ & $<$0.12 & 60.3$^{+0.3}_{-0.2}$ & 18.8\tablenotemark{b} & -40.34$^{+0.05}_{-0.06}$ & $<$9 & 5.102$\pm0.001$\\ 
highres & -0.457$\pm0.002$ & 2.172$^{+0.002}_{-0.001}$ & $<$0.08  & 62.1$\pm0.2$ & 18.8\tablenotemark{b} & -40.37$^{+0.05}_{-0.04}$ & $<$9 & 5.137$\pm0.001$ \\ 
$\delta v_{\rm turb}$=0 & -0.463$\pm0.002$ & 2.168$^{+0.002}_{-0.003}$ & 0\tablenotemark{b} & 62.3$\pm0.1$ & 18.8\tablenotemark{b} & -40.49$^{+0.06}_{-0.05}$ & $<$9 & 5.102$\pm0.001$ \\ 
low $T_{\rm mid0}$ & -0.413$^{+0.001}_{-0.003}$ & 2.208$^{+0.004}_{-0.003}$ & $<$0.18 & 70.02$^{+0.36}_{-0.05}$ & 14\tablenotemark{b} & -40.06$\pm0.05$ & $<$9 & 5.141$\pm0.001$ \\ 
\cutinhead{V4046 Sgr}
Fiducial & -0.565$\pm0.003$ & 1.939$^{+0.004}_{-0.003}$ & $<$0.15 & 27.84$^{+0.13}_{-0.09}$ & 12\tablenotemark{b} & 34.69$^{+0.03}_{-0.04}$ & $<$7 & 2.857$\pm0.001$\\ 
highres & -0.564$^{+0.003}_{-0.001}$ & 1.939$^{+0.001}_{-0.003}$ & $<$0.12  & 28.26$^{+0.10}_{-0.05}$ & 12\tablenotemark{b} & 34.68$^{+0.02}_{-0.03}$ & $<$7 & 2.8957$^{+0.0005}_{-0.0006}$\\ 
$\delta v_{\rm turb}$=0 & -0.579$\pm0.002$ & 1.956$^{+0.004}_{-0.003}$ & 0\tablenotemark{b} & 28.80$^{+0.04}_{-0.07}$ & 12\tablenotemark{b} & 34.76$^{+0.04}_{-0.03}$ & $<$7 & 2.857$^{+0.001}_{-0.001}$\\ 
low $T_{\rm mid0}$ & -0.565$^{+0.002}_{-0.004}$ & 2.078$^{+0.012}_{-0.002}$ & $<$0.18 & 29.27$^{+0.17}_{-0.06}$ & 8\tablenotemark{b} & 34.59$^{+0.05}_{-0.03}$ & $<$6.5 & 2.857$\pm0.001$ \\ 
\enddata
\tablenotetext{}{{\it sys=1.2}: A trial simulating if the data were systematically 20\%\ brighter than measured, {\it sys=0.8}: A trial simulating if the data were systematically 20\%\ fainter than measured, {\it highres}: A trial fitting to the data at full spectral resolution, rather than the binned spectra used in the fiducial model, {\it v$_{turb}$=0}: A trial in which turbulence is fixed at zero, {\it low $T_{mid0}$}: A trial in which the midplane gas temperature is lower than the fiducial model.}
\tablenotetext{a}{$T_{\rm mid0}=19(70/150)^{-q}$. This forces the midplane temperature to 19 K at 70 au, matching the location of the CO condensation front in the chemical modeling of \citet{loo15}}
\tablenotetext{b}{Held fixed during the fitting process.}
\tablenotetext{c}{$T_{\rm mid0}=14(70/150)^{-q}$. }
\end{deluxetable*}

\subsubsection{Anisotropic vs Isotropic motion\label{sec:zonal}}
Within our models we assume that the non-thermal non-Keplerian component of the motion is isotropic turbulence, but other forms of non-Keplerian motion may exist in protoplanetary disks. Deviations are predicted to be induced by planets \citep{per18,bae18} and recent observational studies have found evidence for deviations from Keplerian motion due to the gravitational influence of unseen planets \citep{pin18,tea18b,tea18c,pin19}. Zonal flows, radial density variations associated with magnetic field concentrations, as well as the vertical shear instability (VSI), arising in part from the change in Keplerian velocity with vertical distance from the midplane, produce corrugated variations in the gas motions on velocity scales that are comparable to the non-thermal non-Keplerian motion observed around DM Tau \citep[e.g.][]{sim14,flo17}. If deviations from Keplerian motion are spread through a large region of the outer disk, and are of a sufficiently high velocity at a sufficiently small spatial scale, they may mimic the appearance of isotropic turbulence. 

To test the sensitivity of our data to structured motion, we employ a toy model in which we add a velocity component in the orbital direction that varies as a function of radius:
\begin{equation}
dv_{\phi} = dv\sin\left(\frac{2\pi R}{dR}\right)
\end{equation}
Here $dv$ represents the maximum deviation from Keplerian motion, while $dR$ is the spatial scale of the perturbation. This perturbation is added to a model with parameters ($T_{\rm atm0}$, $R_c$, etc.) based on the DM Tau fiducial model. Visibilities are generated from the model images, and are used to create a cleaned image, using the same procedure as for the data. 

\begin{figure*}
\center
\includegraphics[scale=.33]{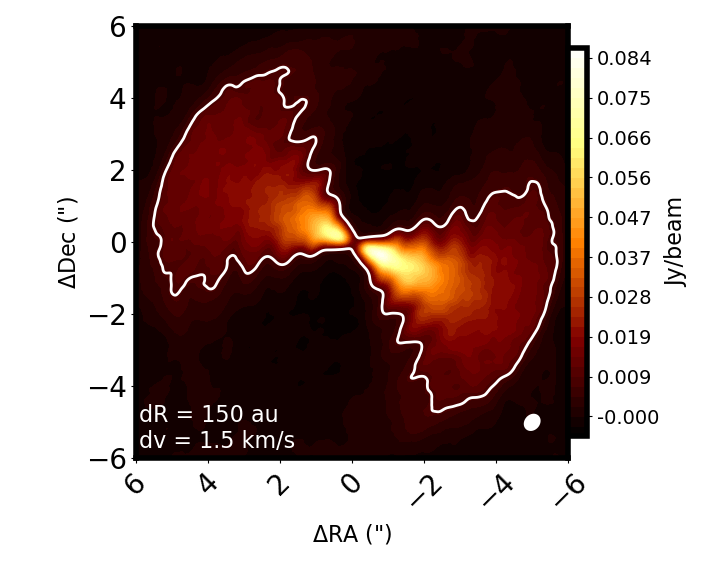}
\includegraphics[scale=.33]{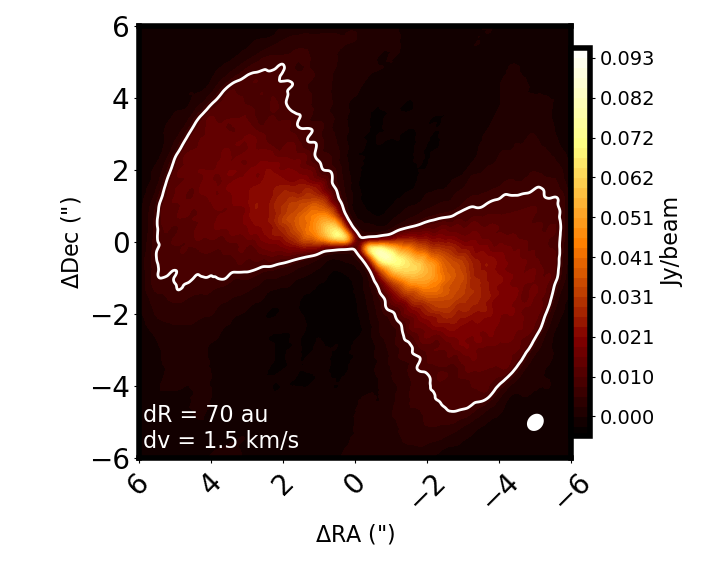}
\includegraphics[scale=.33]{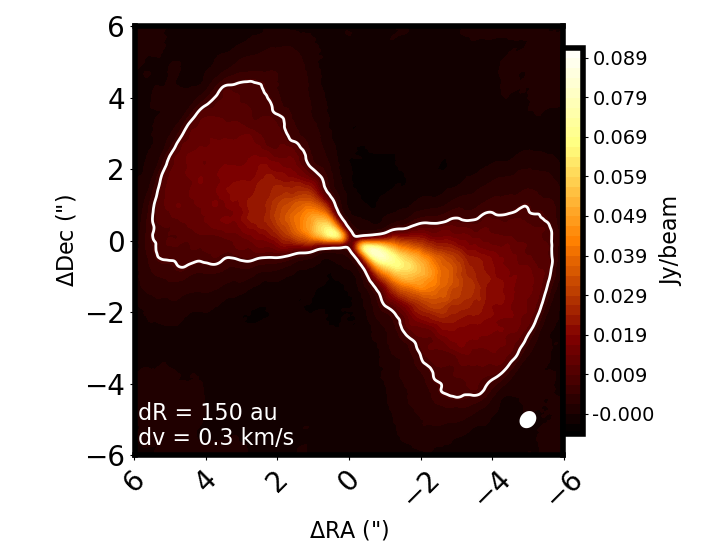}
\caption{Simulated images of the central velocity channel for a radial variation in azimuthal velocity. For features with large velocities and large size scales ($dR$=150 au, $dv$ = 1.5 km s$^{-1}$, left panel), the anisotropic motion displays a corrugated pattern in the images. When the size scale of the features is small ($dR$ = 70 au, center panel) the features are unresolved and appear as a general broadening of the image. When the velocity scale of the features is small ($dv$=0.3 km s$^{-1}\sim$c$_s$, right panel) corrugated features can be difficult to detect even though they are spatially resolved. This indicates that it is difficult for us to distinguish isotropic turbulence from anisotropic motion below the sound speed and at size scales below the resolution of the observations. \label{zonal_vsi}}
\end{figure*}

In Figure~\ref{zonal_vsi} we show the central velocity channel of three models: (1) large spatial scale and high velocity ($dR$=150 au, $dv$=1.5 km s$^{-1}$), (2) small spatial scale and high velocity ($dR$=70 au, $dv$=1.5 km s$^{-1}$), and (3) large spatial scale and small velocity ($dR$=150 au, $dv$=0.3 km s$^{-1}\sim$c$_s$). Anisotropic motion is easily distinguishable from isotropic turbulence in the large size scale and large velocity model due to the corrugated emission structure but becomes more difficult to detect as the size and velocity scales become smaller. When the distance between the minimum and maximum velocity deviation ($dR$/2) becomes smaller than the beam size ($\sim$40 au for DM Tau) the features are no longer spatially resolved, and blend together into a general broadening of the emission. For small velocity scales the azimuthal perturbations to the image are small enough that they are not resolvable as distinct features. Leveraging the full three-dimensional data cube, rather than a single channel as shown in Fig~\ref{zonal_vsi}, provides greater sensitivity to smaller fluctuations \citep{tea18b,tea18c}, although is still limited by the spatial resolution of the data. 

Theoretical models of structured non-Keplerian motion suggest that the radial and velocity scales of typical features are small enough to make them difficult to distinguish from purely isotropic motion in our data. The size scale of these features generally scale with the pressure scale height (H) and for DM Tau our model fits indicate a pressure scale height that reaches 70 au at 500 au from the central star. With a beam size of $\sim$ 40 au, features as large as (5-10)H would be easily resolved. Global non-ideal MHD simulations find that when the net poloidal field is relatively strong (thus driving high accretion rates), the disk tends to form zonal flows of scales\footnote{The cited papers typically discuss the size scale of individual features, e.g. the radial size of a region with higher than Keplerian motion, which corresponds to $dR/2$ in the context of our toy model. In comparing with our sensitivity tests, we quote size scales that are consistent with our definition of $dR$, and hence are typically twice as large as the values mentioned in the cited papers.} of $\sim$2H \citep{bet17,sur18}, while in shearing box simulations, zonal flows are found at a broader range of disk magnetizations, on scales of 4H or beyond and is dependent on the box size \citep{bai14,sim14} and the magnetization of the disk \citep{rio19}. Vertical shear instabilities produce corrugated variations in the gas motion with velocities of 10-20\%\ of the local sound speed and size scales of $\sim$2H \citep{flo17}. At these scales, zonal flows and VSI are difficult to distinguish from isotropic motion in our data, although other factors, such as the efficient lifting of dust grains by VSI \citep{flo17} may break this degeneracy. Planet induced motions are unlikely to be mis-interpreted as turbulence because of the highly localized nature of the perturbation \citep{per18}. In modeling the CO emission from around DM Tau we assume that a single turbulence value, relative to the local sound speed, applies uniformly throughout the disk. The residuals do not show a strong deviation from this assumption, outside of the unresolved emission at the center of the disk. This suggests that instabilities that generate features over a wide radial range within the disk, such as is possible under certain physical conditions with zonal flows and vertical shear instabilities, as well as MRI assuming the surface layers are sufficiently ionized, are consistent with our results around DM Tau.

\subsection{Weak Turbulence around MWC 480 and V4046 Sgr}
\begin{figure*}
\includegraphics[scale=.4]{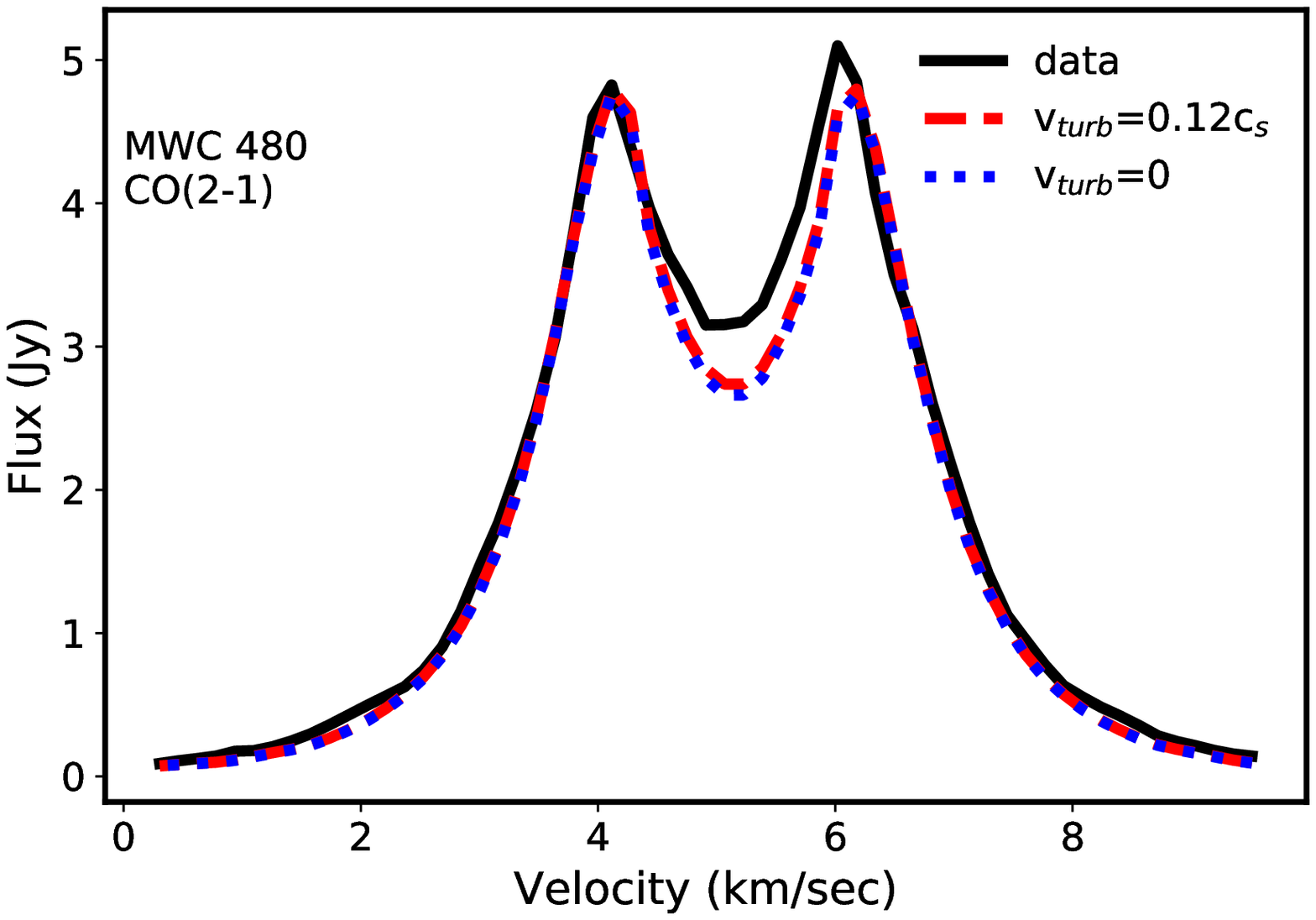}
\includegraphics[scale=.27]{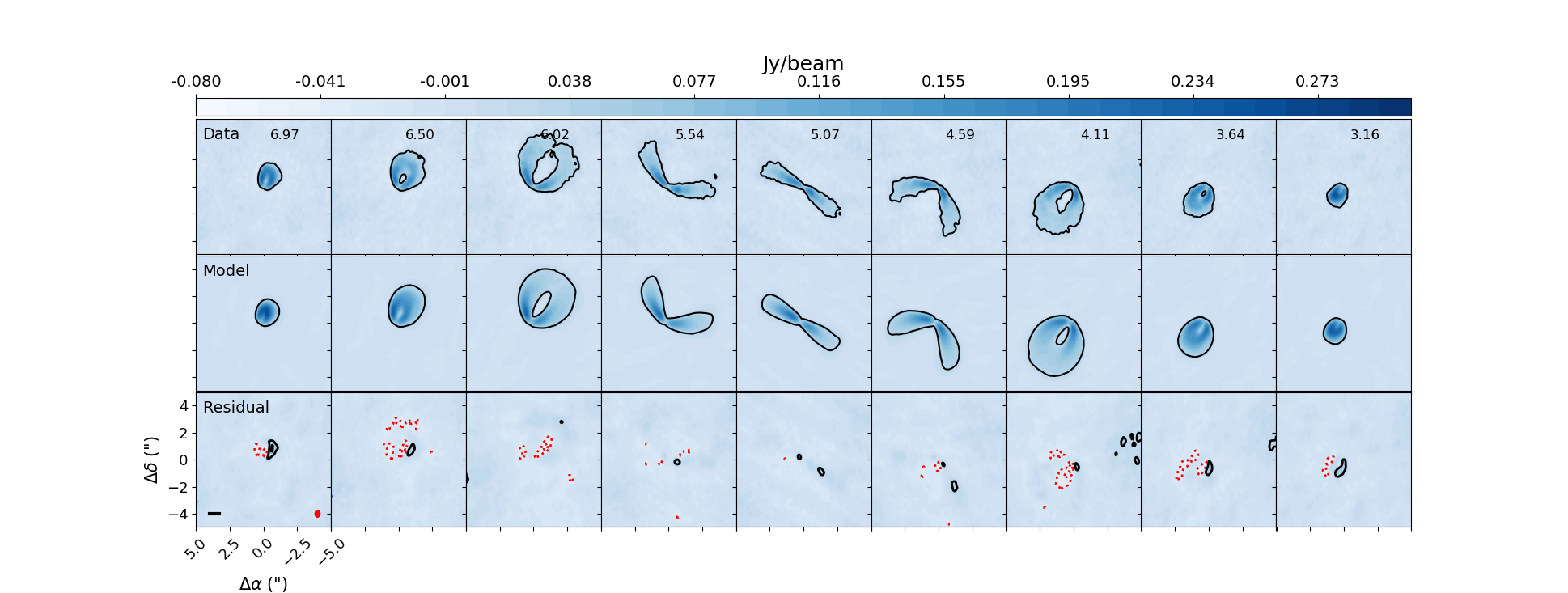}
\caption{(Left:) Spectra of the MWC480 CO(2-1) emission (black line) and the model defined by the median of the PDFs when turbulence is allowed to vary (red dashed line) as well as when turbulence is fixed at zero (blue dotted line).
(Right:) Channel maps of the data (top row), model (middle row), and residuals (bottom row). The contour in the data and model frames is at 5$\sigma$ (=29mJy/beam$\sim$10\%\ of the peak flux), while the contours in the residual frame are at multiples of 5$\sigma$.  \label{mwc480_results}}
\end{figure*}

\begin{figure*}
\includegraphics[scale=.4]{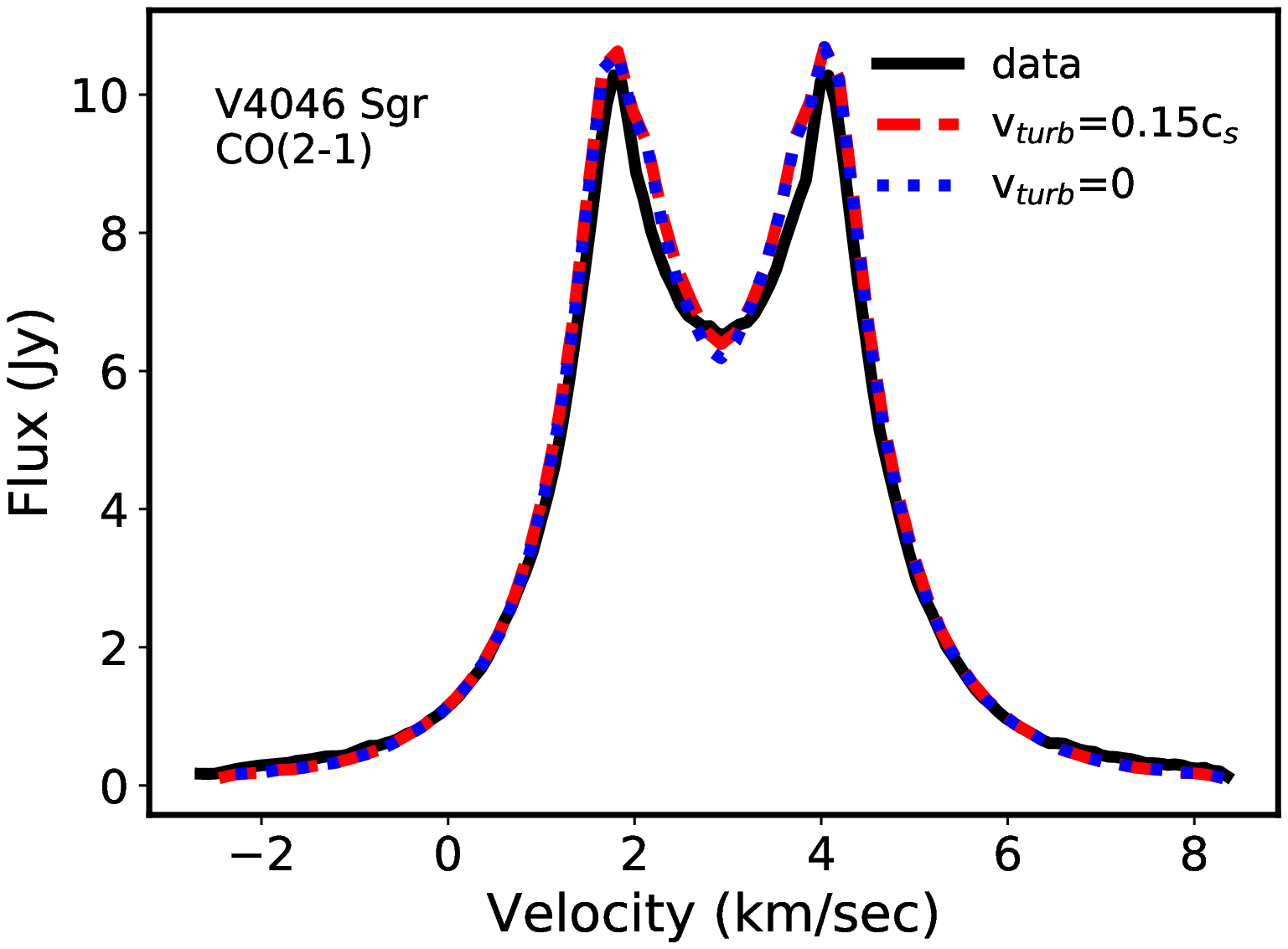}
\includegraphics[scale=.27]{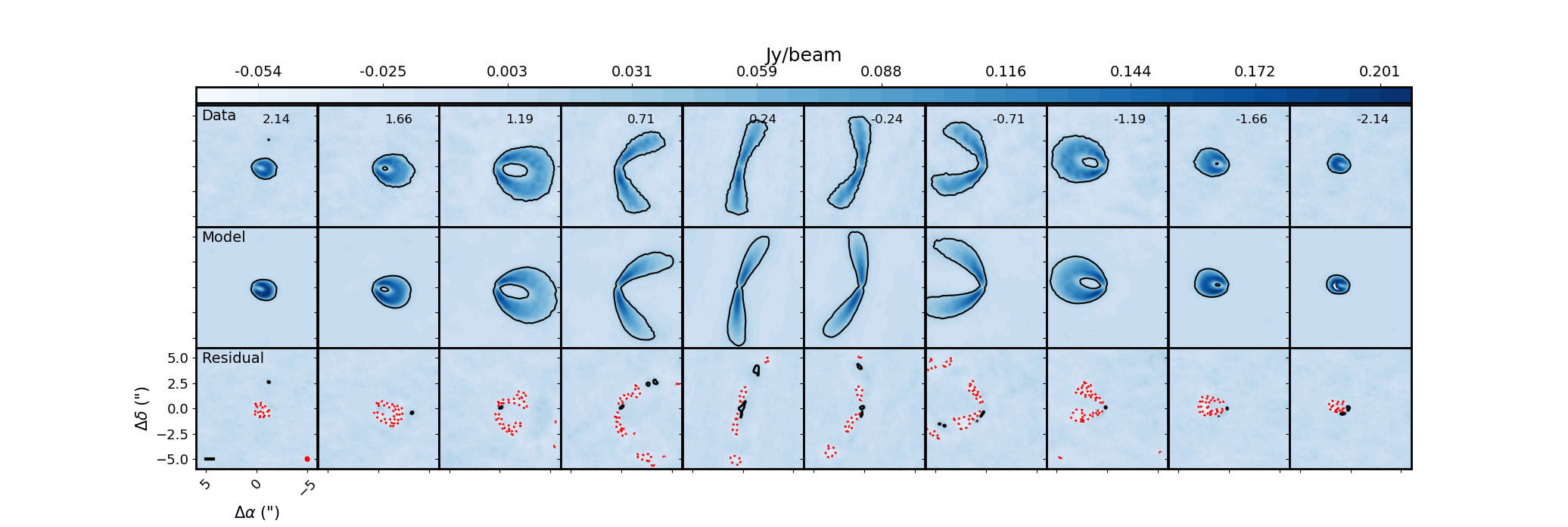}
\caption{(Left:) Spectra of the V4046 CO(2-1) emission (black line) and the model defined by the median of the PDFs when turbulence is allowed to vary (red dashed line) as well as when turbulence is fixed at zero (blue dotted line).
(Right:) Channel maps of the data (top row), model (middle row), and residuals (bottom row). The contour in the data and model frames is at 5$\sigma$ (=24mJy/beam$\sim$7\%\ of the peak flux), while the contours in the residual frame are at multiples of 5$\sigma$.  \label{v4046_results}}
\end{figure*}

For both the disk around MWC 480 and the disk around V4046 Sgr we do not detect significant nonthermal motion of the gas. Figures~\ref{mwc480_results} and ~\ref{v4046_results} show the data as compared to the model defined by the median of the posterior distributions (posterior distributions are shown in Appendix~\ref{pdfs}). For both MWC 480 and V4046 Sgr, the MCMC routine returns a posterior that nominally indicates a detection of turbulence. We run additional trials with turbulence fixed at zero and find that the zero turbulence fit is nearly indistinguishable, in both the spectra and the channel maps, from the non-zero turbulence result, although as discussed below there are still residuals between the best fit models and the data. For this reason we conservatively interpret the results for MWC 480 and V4046 Sgr as upper limits rather than detections. 


For MWC 480, we are able to constrain turbulence to $<$0.12c$_s$. Fitting to the un-binned data pushes the limit down to $<$0.08 c$_s$, corresponding to velocities below 20-40 m s$^{-1}$ beyond 100 au. For V4046 Sgr a fit to the binned spectrum constrains turbulence to $<$0.15c$_s$, while a fit to the un-binned spectrum pushes the upper limit down to $<$0.12c$_s$, corresponding to velocities below 30-55 m s$^{-1}$ beyond 100 au. As with DM Tau, we consider models with colder midplane temperatures ($T_{\rm mid0}$ = 14 K for MWC 480, and 8 K for V4046 Sgr) and in both cases the upper limits rise to $<$0.18 c$_s$, but are still consistent with a non-detection of turbulence. This suggests that systematic effects (e.g., the assumed midplane temperature) are significant compared to statistical uncertainties when constraining the turbulence level in these systems. 

In the case of both MWC 480 and V4046 Sgr, the imaged residuals show structure (Figure~\ref{mwc480_results}, \ref{v4046_results}), although at a level that is less than 10\%\ of the peak flux. Both systems show an unresolved positive residual at the center of the disk, with a diameter $<$50 au and $<$20 au for MWC 480 and V4046 Sgr respectively, which may be a result of a deviation from simple prescriptions for the temperature and surface density radial profiles, a change in CO abundance with radius \citep[e.g.][]{sch18}, or a change in the velocity profile due to a warp \citep[e.g.][]{wal17} or rapid inward radial flow \citep[e.g.][]{ros14}. Around V4046 Sgr negative residuals extend out to $\sim$200 au, corresponding to the radius beyond which our simple photo-desorption treatment sets the CO abundance. This suggests that a more detailed treatment of photodesorption would bring the model in closer agreement with the data. None of these residuals are indicative of an underestimate of the turbulence level. In the channel maps turbulence acts to broaden emission rather than to change the radial profile \citep{sim15}. Also, the similarity between the model derived when turbulence was fixed at zero versus the model derived when turbulence was allowed to vary supports our interpretation of the model results as an upper limit. We can rule out turbulence levels comparable to that around DM Tau, but cannot completely rule out modest turbulent motions within these systems.

\section{Discussion\label{discussion}}
\subsection{Is weak turbulence common among planet-forming disks?}
Based on an analysis of CO(2-1) ALMA observations of the disks around DM Tau, MWC 480, and V4046 Sgr we find significant non-zero turbulence around DM Tau, but only upper limits for the disks around MWC 480 and V4046 Sgr, similar to what has been found in the disks around HD 163296 \citep{fla15,fla17} and TW Hya \citep{fla18,tea18a}. 

Molecular line emission provides a direct measure of gas motions, but it is not the only tracer of turbulence in the outer disk. Strong turbulence will loft dust grains away from the disk midplane \citep{you07,tun10}, making dust settling an indirect tracer of turbulence, in particular the turbulence at the midplane of the disk. Using the influence of dust settling on the SED, \citet{bon16} find $\alpha\sim(0.1-6.3)\times10^{-3}$ around HD 163296, while \citet{mul12} use the average SEDs of T Tauri stars, Herbig stars, and brown dwarfs and find $\alpha$=10$^{-4}$ is a better fit to the data than $\alpha$=10$^{-2}$, for regions of the disk within $\sim$100 au of the central star. In combining scattered light observations and the SED, \citet{van17} find $\alpha\sim2\times10^{-4}$ around TW Hya. \citet{qi19} find that, in SED fitting and comparison with the N$_2$H$^+$ emission morphology, the disk around V4046 Sgr exhibits more dust settling, and hence has weaker turbulence, than the disk around DM Tau, consistent with our results.

High resolution images of gaps in planet-forming disks have opened up additional probes of turbulence through the effect of turbulence on the appearance of dust and gas gaps. The well defined dust gaps around HL Tau are a sign of substantial dust settling, as expected for $\alpha\sim10^{-4}$ \citep{pin16}. Using hydrodynamic simulations of gas and dust gap formation via gravitational interactions with massive planets, \citet{liu18} find that $\alpha$ increases with radius from $<$10$^{-4}$ to 8$\times10^{-3}$ in the disk around HD 163296, while \citet{oha19} find that the polarization pattern is also consistent with a rise in $\alpha$, although at a factor of ten higher level ($\alpha\lesssim1.5\times10^{-3}$ at 50 au rising to 0.015-0.3 at 90 au). In high resolution ALMA images the disks around AS 209, DoAr 25, Elias 20, and RU Lup \citep{hua18} and HD 169142 \citep{per19} exhibit `double gap' features consistent with clearing by a single planet in a low viscosity disk \citep[$\alpha<10^{-4}$][]{don17}. Overall, while observations are limited to the brightest systems, they suggest that large turbulence is not a frequent occurrence around protoplanetary disks at least in the mid-plane.

In the context of the $\alpha$-disk model, turbulent velocities from molecular line observations can be converted to $\alpha$ via $\alpha\sim$($\delta$v$_{\rm turb}$/c$_s$)$^2$, but caution is needed when comparing to $\alpha$ values derived in different observational studies, as well as $\alpha$ values from numerical simulations. Direct conversion from turbulent velocities to $\alpha$ values results in $\alpha$=0.078$\pm$0.002 for the disk around DM Tau, $\alpha<0.006$ for disk around MWC 480 and $\alpha<$0.01 for the disk around V4046 Sgr. These $\alpha$ values only apply to the localized region where the CO emission arises, which is well above the midplane (e.g. Figure~\ref{turb_struct_dmtau}). Observations of millimeter dust continuum emission will be more sensitive to the turbulence at smaller radii than the gas, since dust emission is often more compact than gas emission \citep[e.g.][]{kas18}, and at the midplane, in part because large grains quickly settle to the midplane and because the scale height of the dust is most sensitive to the turbulence near the midplane \citep{cie07}. If the turbulence varies with vertical height, as is expected for MRI driven turbulence, the local $\alpha$ value will change with height as well. For example, in an MRI simulation at 100 au from \citet{sim15} with an initial gas to magnetic pressure ratio at the midplane of $\beta_0$=10$^5$ and an FUV penetration depth of $\Sigma_{\rm FUV}$=0.01 g cm$^{-2}$ the turbulent velocity varies with height by a factor of $\sim$10 (0.7c$_s$ at 3 pressure scale heights vs. 0.04c$_s$ at the midplane), resulting in a factor of $\sim$100 difference between the local $\alpha$ values ($\alpha$=0.49 at 3 pressure scale heights vs. 0.002 at the midplane). These local $\alpha$ values are distinct still from the density-weighted average of the stress tensor ($\alpha$=0.0026), which is related to the accretion flow through the disk and is weighted more strongly toward the high density midplane. Measurements of the turbulence near the midplane through optically thin molecular line tracers or dust observations, in addition to confirmation that the measured non-thermal motion around DM Tau is due to turbulence and not unresolved structured motion, are needed to fully characterize the $\alpha$-viscosity in these systems.


\begin{deluxetable*}{cccccccc}
\tabletypesize{\scriptsize}
\tablewidth{0pt}
\tablecaption{System properties\label{properties}}
\tablehead{\colhead{Star} & \colhead{$\delta v_{\rm turb}$} & \colhead{M$_{*}$ } & \colhead{M$_{\rm disk}$ } & \colhead{Age} & \colhead{$\dot{M}_{\rm acc}$ } & \colhead{L$_{\rm X-ray}$ } & \colhead{L$_{\rm FUV}$ } \\ 
\colhead{} & \colhead{} & \colhead{M$_{\odot}$} & \colhead{M$_{\odot}$} & \colhead{Myr} & \colhead{M$_{\odot}$ yr$^{-1}$} & \colhead{L$_{\odot}$} & \colhead{L$_{\odot}$}}
\startdata
HD 163296 & $<$0.05c$_s$ (1) & 2.3 & 0.09 (2)  & 5 & 5$\times$10$^{-7}$ (3) & 10$^{-4}$  (4, 5) & 3.21-5.58 (6)\\
TW Hya & $<$0.08c$_s$ (7) & 0.6 & 0.05 (8) & 10-12 & 2$\times$10$^{-9}$ (9, 10, 11) & 6.8$\times$10$^{-4}$ (12, 13, 14, 15) & 7$\times$10$^{-3}$ (16)\\
MWC 480 & $<$0.08c$_s$ (17) & 1.85 (18) & 0.046 (17) & 7-8 & 5.3$\times$10$^{-7}$ (19, 3, 20) & 2.6$\times$10$^{-6}$ (21) & $\ldots$\\
V4046 Sgr & $<$0.12c$_s$ (17) & 0.9,0.9 & 0.09 (22) & 12-23 (23, 24, 25) & 1.3$\times$10$^{-8}$ (26) & 3.1$\times$10$^{-4}$ (27, 28) & 10$^{-2}$ (16)\\
DM Tau & 0.25-0.33c$_s$ (17) & 0.54 (29) & 0.04 (30,31) & 1-5 & 2.9$\times$10$^{-9}$  (16, 32) & 7.8$\times$10$^{-5}$ (33) & 3$\times$10$^{-3}$ (34, 16)\\
\hline
Typical Disk Ranges& $\ldots$ & $\ldots$ & 3$\times$10$^{-4}$ - 0.03 (35) & 1-10 & 5$\times$10$^{-11}$ - 10$^{-7}$ (36) & 2$\times10^{-6}$ - 0.02 (37) & 2$\times10^{-3}$ - 0.2 (16)\\
\enddata
\tablenotetext{}{References: (1) \citet{fla17} (2) \citet{ise07}, (3) \citet{men13}, (4) \citet{gun09}, (5) \citet{scw05}, (6) \citet{mee12}, (7) \citet{fla18}, (8) \citet{ber13}, (9) \citet{ale02}, (10) \citet{her04}, (11) \citet{ing13}, (12) \citet{bri10}, (13) \citet{dup12}, (14) \citet{kas99}, (15) \citet{kas02}, (16) \citet{fra14}, (17) This work, (18) \citet{pie07} (19) \citet{cos14}, (20) \citet{don11}, (21) \citet{hen10}, (22) \citet{ros13b}, (23) \citet{mam14} (24) \citet{tor06}, (25) \citet{bin14}, (26) \citet{cur11}, (27) \citet{esp13}, (28) \citet{sac12}, (29) \citet{sim00}, (30) \citet{and11}, (31) \citet{mcc16}, (32) \citet{ing11}, (33) Cleeves, private communication, (34) \citet{yan12}, (35) \citet{and13}, (36) \citet{man16}, (37) \citet{fei05}  }
\end{deluxetable*}

\subsection{Physical conditions for turbulence}
Physical conditions within the disk can set the turbulence level and may explain the lack of turbulence among many of the studied sources. In the context of MRI, \cite{sim18} explore the turbulence levels among models with different ionization conditions and magnetic field strengths. They find that the presence of FUV emission drives strong turbulence (0.2c$_s$ - 0.7c$_s$), while the removal of this ionization component, especially when combined with weak magnetic fields, can lead to weak turbulence (0.02c$_s$ - 0.09c$_s$). Among the sources with constraints on turbulence from molecular line emission, DM Tau exhibits the weakest FUV flux (Table~\ref{properties}) despite having the only turbulent disk. X-ray emission can also ionize gas, but again DM Tau does not posses the strongest X-ray emission among our sample. This suggests that FUV and X-ray emission strength alone does not set the turbulence level.  

Observations constrain the FUV and X-ray emission near the stellar surface, while the relevant parameter for MRI models is the FUV  and X-ray field as seen by material in the outer disk. High energy radiation can be severely attenuated before reaching the outer disk if intervening material, such as a wind launched by the inner disk, acts as an obscuring screen. TW Hya \citep{pas11}, MWC 480 \citep{fer18}, V4046 Sgr \citep{sac12}, and HD 163296 \citep{kla13} show evidence of a disk wind based on optical/infrared forbidden lines (TW Hya, V4046 Sgr), infrared emission (MWC 480), and extended CO emission (HD 163296). In contrast, the optical forbidden line emission from DM Tau can be explained entirely with a gas disk in Keplerian motion \citep{sim16}. Whether or not a wind is the defining factor between a turbulent and non-turbulent disk depends on the density of the wind, with vertical column densities of 0.01 g cm$^{-2}$ sufficient to absorb a majority of the FUV radiation \citep{per11}.

Regardless of the ionizing source, the ionization level of the outer disk can be constrained based on observations of ionized molecular species. The presence of e.g. DCO$^+$, HCO$^+$, N$_2$H$^+$ in the outer disk of HD 163296 \citep{fla17}, TW Hya \citep{qi13}, DM Tau \citep{dut97,obe11,tea15}, MWC 480 \citep{obe10,hua17} and V4046 Sgr \citep{obe11b,hua17} suggests that some ionizing radiation reaches the outer disk. \citet{obe11} measure an ionization fraction of $\sim$4$\times10^{-10}$ in the CO layer around DM Tau, comparable to the ionization fraction in the models from \citet{sim18} that assumed no FUV photons but still included X-rays and cosmic-rays. A subset of these models displayed vigorous turbulence in the upper disk layers.

\citet{cle17} found that the H$^{13}$CO$^+$ abundance around IM Lup varied in response to an X-ray flare, indicating that the ionization level may change on short timescales. In the presence of variable ionization, MRI will respond with varying turbulence levels on its growth/decay timescale. Since the MRI relies on the differential rotation of gas, its growth/decay timescale is, at its fastest, the dynamical timescale of the gas. Molecular line observations with ALMA probe regions of the disk beyond $\sim$30 au, corresponding to dynamical timescales of hundreds to thousands of year. In their chemical modeling, \citet{cle17} find that the HCO$^+$ abundance would be enhanced for 10-20 days at 100 au following an X-ray flare, much shorter than the time needed for MRI to respond to changes in the ionization, resulting in little variation in turbulence with time. Longer timescale fluctuations in e.g. the wind mass loss rate may still lead to oscillations in the turbulence level over the course of the disk lifetime. 

For reasonable values of the magnetic field strength, as informed by theoretical models \citep[e.g.][]{bai11,bai16,sim18}, stronger fields generally result in stronger turbulence, and possibly higher accretion rates. The predicted correlation between turbulence and accretion rate seems to be contradicted in our sample, with DM Tau showing the strongest turbulence at yet the weakest accretion rate. This may be due to the fact that the measured accretion rate is the instantaneous accretion rate onto the star, and does not reflect the accretion rate in the outer disk where the turbulence is measured, or that the turbulence signature does not reflect MRI-generated turbulence but instead reflects other sources of non-thermal motion (e.g., unresolved spatial variations in azimuthal velocity).

Age may also play a factor in the strength of the turbulence by driving the evolution of the physical conditions that set the turbulence level. DM Tau is one of the youngest targets within our sample; as a member of the Taurus star forming region its age likely lies between 1 Myr \citep{luh10} and 5 Myr \citep{sim00,gui14}. Larger turbulence at younger ages is consistent with the findings of \citet{naj18} of large $\alpha$ values among class I objects. But ages of individual objects are difficult to constrain; DM Tau and MWC 480 are both part of the Taurus region, and would be expected to have similar ages, yet MWC 480 is typically assumed to be older than DM Tau \citep{sim00}. Accounting for internal magnetic fields can push stellar ages to systematically higher values \citep{sim19}. Similarly, the lack of turbulence around the class I object HL Tau \citep{pin16} suggests that not all objects of the same age have the same turbulence properties.

In the absence of MRI, other processes may drive turbulence under the right conditions. \citet{for12} find that gravitational instabilities can drive large scale accretion, with the driving source transitioning from large scale spiral arms to gravito-turbulence as the disk mass decreases. The measured disk masses (Table~\ref{properties}) are 4-8\%\ of the stellar mass, resulting in minimum Toomre Q values of 5-15, inconsistent with the presence of gravito-turbulence. \citet{pow19} derive larger disk masses for HD 163296 and TW Hya based on the structure of the dust emission, but they find that even these large masses do not push these disks into the gravitationally unstable regime. The weak levels of turbulence may also open up the possibility of hydrodynamic instabilities, which can produce $\alpha\sim10^{-4}-10^{-3}$, playing a larger role  \citep[e.g. Vertical Shear Instability, Convective Overstability, and Zombie Vortex Instability as reviewed by ][]{lyr19}. The Vertical Shear Instability (VSI) develops due to the change in orbital velocity with vertical height within a disk that has sufficiently rapid cooling \citep{sim14,flo17}, conditions that are likely in the outer disk \citep{lin15}. VSI can operate in a magnetized disk \citep{cui19}, and produces highly anisotropic motion, dominated by vertical motion reaching 10-20\%\ of the local sound speed \citep{flo17}. Such motions can be ruled out for some of the sample HD 163296, TW Hya, MWC 480). Relying on the ability of turbulence to spread dust in the radial direction, \citet{dul18} use the resolved widths of narrow dust rings around AS 209, Elias 24, HD 163296, GW Lup, and HD 143006 to rule out $\alpha\ll$5$\times10^{-4}$ for grain sizes $\gg$0.1 cm, suggesting that some modest level of turbulence may exist in these systems.

\section{Conclusion}
Using new ALMA observations of CO(2-1) emission, we detect turbulence around DM Tau at a level of $\delta v_{\rm turb}=$0.279$^{+0.005}_{-0.004}$c$_s$, although this range extends from 0.25c$_s$ to 0.33c$_s$ when accounting for systematic effects (e.g., assumptions about the midplane temperature). Around MWC 480 and V4046 Sgr we find no strong evidence for non-zero turbulence, restricting the turbulence levels to $<$0.08c$_s$ and $<$0.12c$_s$ respectively. The finding of larger turbulence around DM Tau than around V4046 Sgr is consistent with the enhanced dust settling in the disk around V4046 Sgr relative to the disk around DM Tau \citep{qi19}.


The upper limits on turbulence in the disks around MWC 480 and V4046 Sgr put them at a turbulence level similar to that around HD 163296 \citep{bon16,fla17,liu18,dul18} and TW Hya \citep{van17,fla18,tea18a}. While this is still a modest sample, the results are consistent with studies of high resolution images of continuum emission, sensitive to disk properties closer to the midplane and smaller radii than the CO emission, which found modest turbulence among additional planet-forming systems \citep[$\alpha\sim10^{-4}$,][]{pin16,hua18,per19}, suggesting that weak turbulence may be a common feature among planet-forming disks. At the same time the robust detection of turbulence around DM Tau provides an important anchor point in comparison with models of instabilities within planet-forming disks. 


Amongst this small sample we can begin to look for clues as to the explanation for these results. In the context of MRI, the strength of the ionizing radiation reaching the outer disk and the magnetic field are two defining factors \citep[e.g.][]{sim18}. While the FUV and X-ray flux emitted by DM Tau is not large relative to MWC 480, V4046 Sgr, HD 163296, or TW Hya, DM Tau is the only source among this sample that does not show evidence for an inner disk wind that could potentially block the ionizing radiation from reaching the outer disk. The presence of ionized molecular species in all of these systems indicate that some modest level of ionization exists in the outer disk, although the exact ionization level has only been quantified in the case of the disk around DM Tau. Instead the magnetic field strength may be the defining factor in setting the strength of the turbulence. Magnetic field strengths remain largely unconstrained, with observational upper limits \citep[e.g.][]{vle19} still lying above the levels examined by theoretical models. Age also may play a role, given that DM Tau is the youngest source among our sample, although further measurements are needed to determine if this trend holds up.






\acknowledgments
The work of KMF was performed in part at the Aspen Center for Physics, which is supported by the National Science Foundation grant PHY-1607611. A.M.H. is supported by a Cottrell Scholar Award from the Research Corporation for Science Advancement. This paper makes use of the following ALMA data: ADS/JAO.ALMA\#2016.1.00724.S. ALMA is a partnership of ESO (representing its member states), NSF (USA) and NINS (Japan), together with NRC (Canada), MOST and ASIAA (Taiwan), and KASI (Republic of Korea), in cooperation with the Republic of Chile. The Joint ALMA Observatory is operated by ESO, AUI/NRAO and NAOJ. The National Radio Astronomy Observatory is a facility of the National Science Foundation operated under cooperative agreement by Associated Universities, Inc. This project has received funding from the European Research Council (ERC) under the European Union's Horizon 2020 research and innovation programme under grant agreement No 716155 (SACCRED).

\vspace{5mm}
\facilities{ALMA}


\software{astropy \citep{astropy}, EMCEE \citep{for13}, MIRIAD}


\appendix

\section{CO photo-desorption\label{CO_photo_desorption}}

To incorporate CO in regions with photo-desorption into our parametric structure, without resorting to detailed chemical models, we compare the timescales for freeze-out, photo-desorption and photodissociation. Regions in which photo-desorption operates more quickly than freeze-out and photo-dissociation will provide additional CO emission in the outer disk. Regions in which freeze-out or photo-dissociation dominate will be heavily depleted in CO, and will not contribute CO emission. 

The freeze-out timescale is set by the frequency at which CO molecules collide with dust particles \citep{flo05}.
\begin{equation}
\tau_{fre} = (n_g \pi a_g v_{th} S)^{-1},
\end{equation}
where $n_g$ is the number density of grains, $a_g$ is the radius of a grain particle, $v_{\rm th}$ is the thermal velocity of the gas, and $S$ is the sticking efficiency, assumed to be 1. 

CO photo-desorption is influenced by the UV radiation field, its ability to penetrate the disk, and the intrinsic efficiency with which CO desorbs from a dust grain in response to an incoming photon:
\begin{equation}
R = I_{ISRF} \exp{(-\gamma A_V)} Y_{pd} a_g
\end{equation}
where $R$ is the number of CO molecules that are photo-desorbed per second per dust grain, $I_{\rm ISRF}$ is the interstellar radiation field \citep[$=$1 Habing$=10^8$\,photons\,s$^{-1}$\,cm$^{-2}$][]{hab68}, $\gamma$ is a measure of the UV extinction relative to the visual extinction ($\sim$2 for small dust grains), $A_V$ is the visual extinction, and $Y_{\rm pd}$ is the experimentally measured number of photo-desorptions per UV photon. A detailed calculation of the interstellar radiation field would involve accounting for the nearby stellar population \citep[e.g.][]{cle16}, while here we assume a radiation field of 1 Habing for simplicity. The factor $Y_{\rm pd}$ can vary by $\sim$3-4 depending on the exact shape of the incident spectrum \citep{fay11,che14} and the temperature at which the ice is deposited on the dust grains \citep{obe09}, although we ignore these effects in our calculation, and assume $Y_{\rm pd}$=2.7$\times10^{-3}$. Given the photo-desorption rate ($R$), the timescale to double the CO density is:
\begin{equation}
\tau_{pde} = \frac{n_{CO}}{R n_g}
\end{equation}
where $n_{\rm CO}$ is the number density of CO molecules.

Once CO is released from the grains it must survive as a molecule, without being dissociated by the same UV radiation field that released it from the dust grain in the first place. The rate at which CO is photo-dissociated is \citep{vis09}:
\begin{equation}
k = \chi k_0 \Theta \exp{(-\gamma A_V)}
\end{equation}
where $\chi$ is the radiation field in units of the \citet{dra78} field (which corresponds to roughly 1.7 Habings), $k_0$ is the unattenuated photo-dissociation rate (=$2.592\times10^{-10}$, in units of photo-dissociations per second per CO molecule), and $\Theta$ is the attenuation factor accounting for shielding of the radiation field by H$_2$ as well as self-shielding by CO. The photo-dissociation timescale is simply $\tau_{pdi}=1/k$. 

\begin{figure*}
\center
\includegraphics[scale=.6]{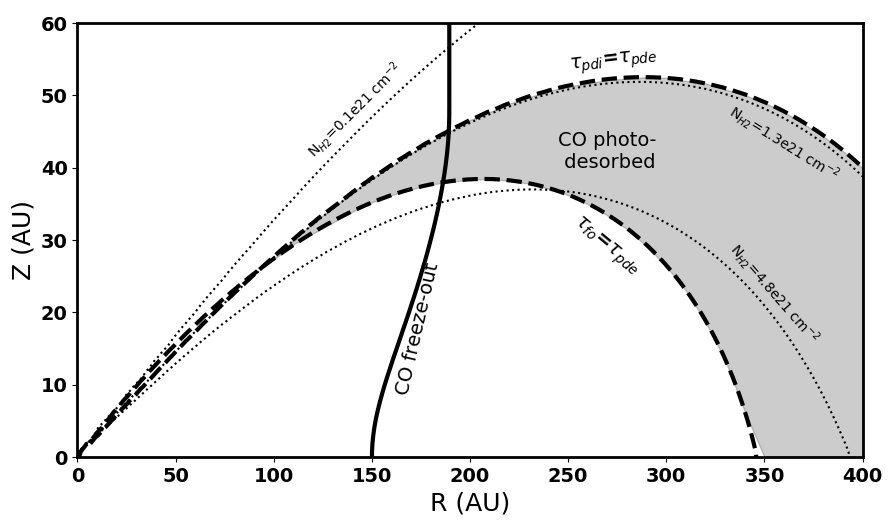}
\includegraphics[scale=.4]{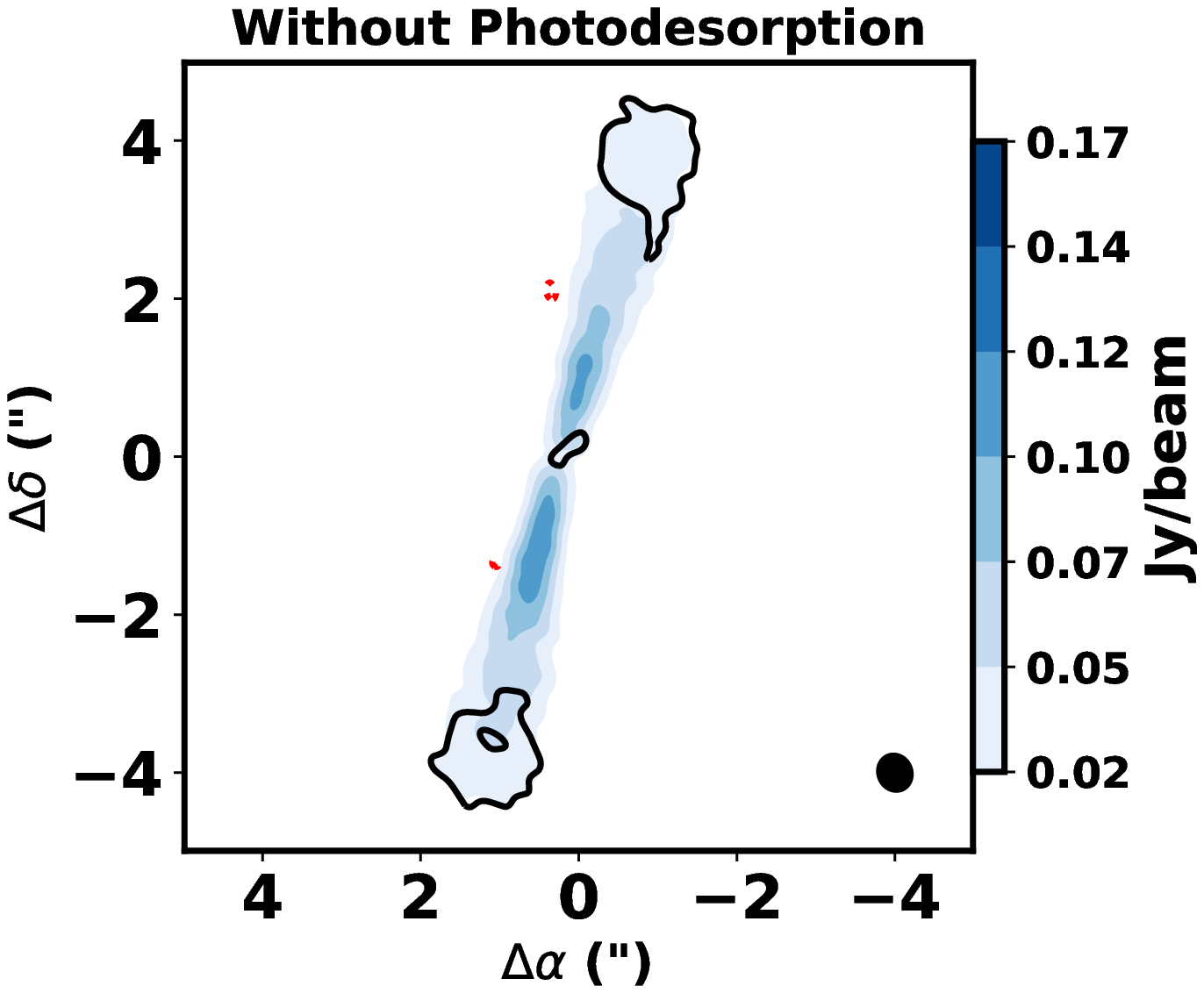}
\includegraphics[scale=.4]{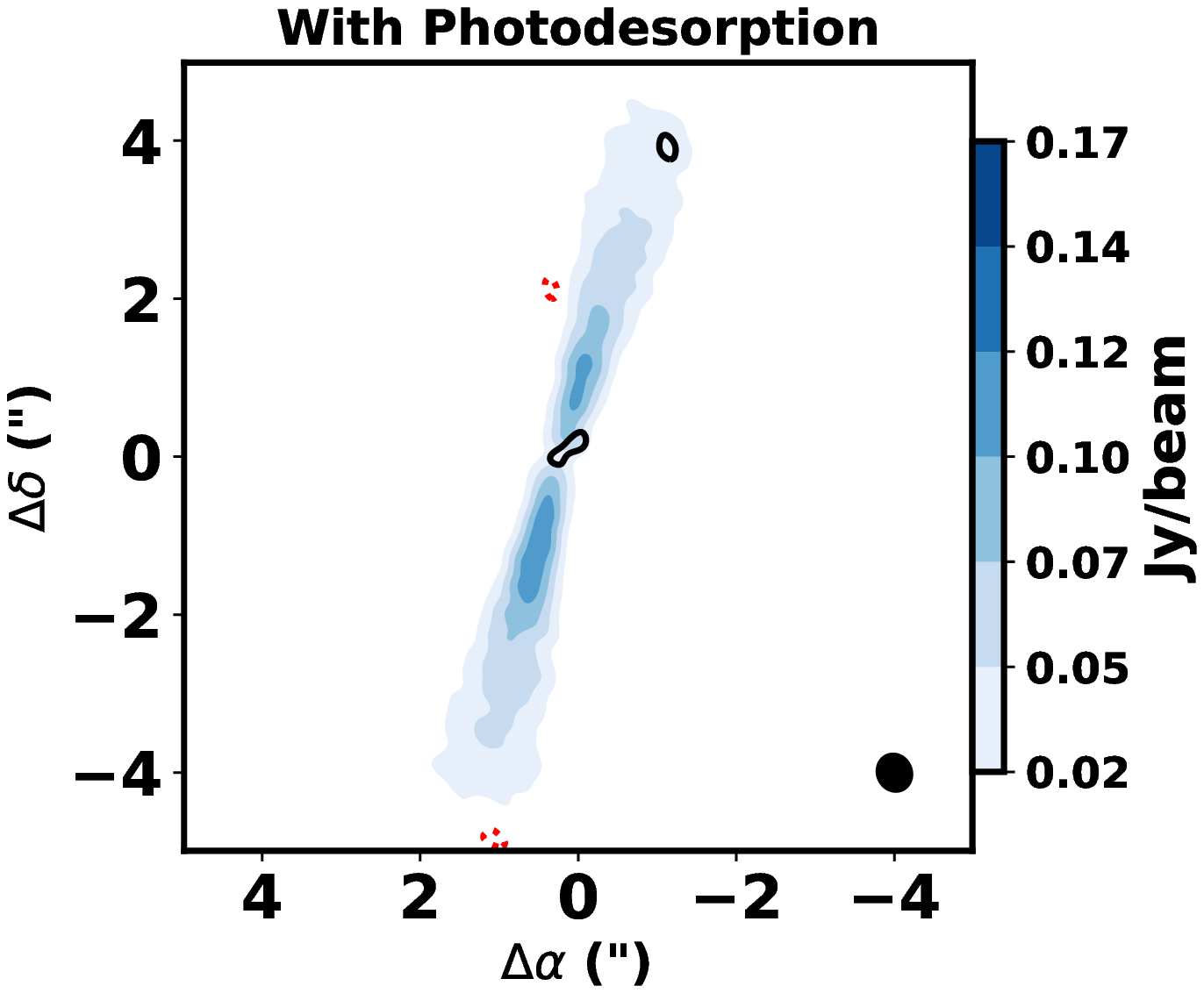}
\caption{({\it Top: }) Disk profile, indicating the boundaries between which photo-desorption can return CO from the solid phase to the gas phase. Near the midplane the freeze-out timescale is smaller than the photo-desorption timescale, due to the increase grain density and weaker radiation field, while in the surface layers photo-dissociation operates more quickly than photo-desorption, due to the limited influence of self-shielding. The boundaries where the freeze-out timescale equals the photo-desorption timescale, and where the photo-dissociation timescale is equal to the photo-desorption timescale are marked by the lower and upper dashed lines respectively. Between these two boundaries photo-desorption can return CO to the gas phase. The boundaries of this region closely follow contours of constant vertically integrated surface density (marked with dotted lines). 
({\it Bottom:}) Residuals (black contours for data$>$model, red dotted contours for model$>$data) between the data (shown in the blue-filled contours) and a model without photodesorption (left) and including our prescription for photodesorption (right) for the central velocity channel of the CO emission from around V4046 Sgr. The addition of photodesorption accurately reproduces the majority of the emission in the outer disk that would otherwise not be included in the model.  \label{pde}}
\end{figure*}

Using the disk around V4046 Sgr as a reference, with the parameters taken from \citep{ros13b} and assuming 0.1$\micron$ dust grains and a dust-to-gas-ratio of 10$^{-3}$, we find a region in the upper layers of the outer disk in which photo-desorption operates more quickly than freeze-out and photo-dissociation (Fig~\ref{pde}). Near the disk surface photo-desorption operates more quickly than freeze-out due to the strong UV field and the low densities. In the uppermost layers photo-dissociation operates more rapidly than photo-desorption; while both processes rely on the same UV field, the decreasing number of dust grains from which CO can desorb, as well as the diminishing effects of self-shielding, lead photo-dissociation to operate more quickly than photo-desorption in the highest layers. These trends limit effective photo-desorption to a narrow layer within the disk, similar to what has been seen in more detailed chemical modeling \citep{obe15,tea17}. The boundaries for the photo-desorption region appear to follow contours in the vertically integrated column density. Given this behavior we parameterize CO non-thermal desorption as a return to normal CO abundance in the region between $N_{H_2}=1.3$ and 4.8 $\times$10$^{21}$ cm$^{-2}$. The CO emission is most strongly dependent on the boundary at lower column densities, corresponding to larger heights above the midplane, due to the highly optically thick nature of CO. Our choice of $N_{H_2}=1.3\times$10$^{21}$ cm$^{-2}$ for photodesorption matches the boundary used in the inner disk for photodissociation of CO, ensuring that there is no discontinuity in the CO location across the freeze-out boundary. Including this parameterization for CO photo-desorption eliminates most of the residuals in the outer disk (Fig~\ref{pde}).

\section{Positive vs Negative Inclination\label{inclination_sign}}
For a perfectly thin disk, there is a degeneracy between positive and negative inclinations, but protoplanetary disks have a nonzero thickness, which introduces small differences in the emission based on the sign of the inclination. In the case of the disk around HD 163296, \citet{ros13} demonstrated that spatially resolving the upper and lower molecular layers of the disk indicates that the inclination is positive. While we do not resolve the layers of the disks around DM Tau, MWC 480, or V4046 Sgr, the nonzero thickness still influences the emission. Lines of sight through the half of the disk closer to the observer terminate at smaller radii, and hence warmer gas temperatures, than lines of sight through the far side of the disk. This leads to a small asymmetry in the emission from the disk that can be detected when the data are of sufficiently high S/N. 

\begin{figure*}
\center
\includegraphics[scale=.4]{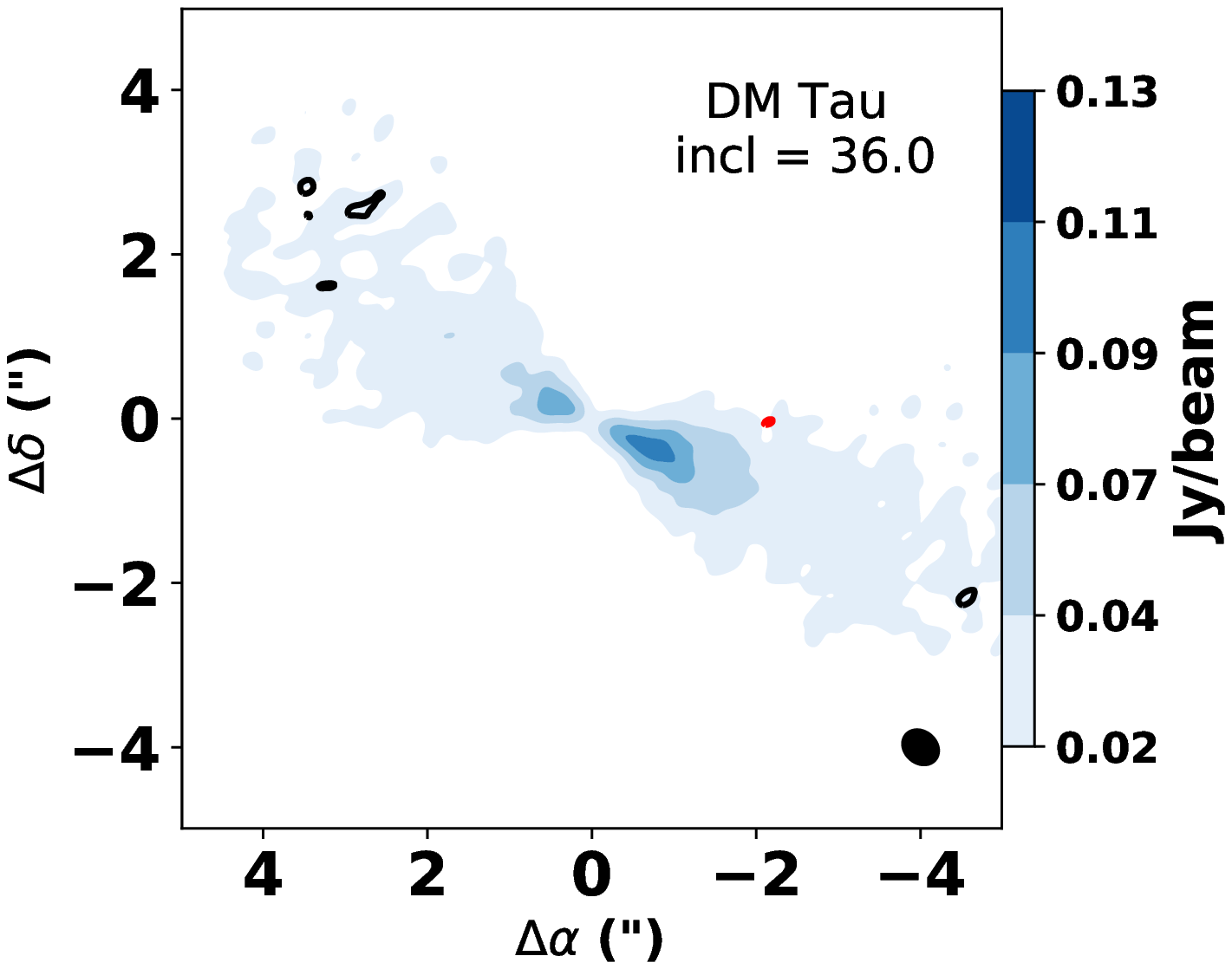}
\includegraphics[scale=.4]{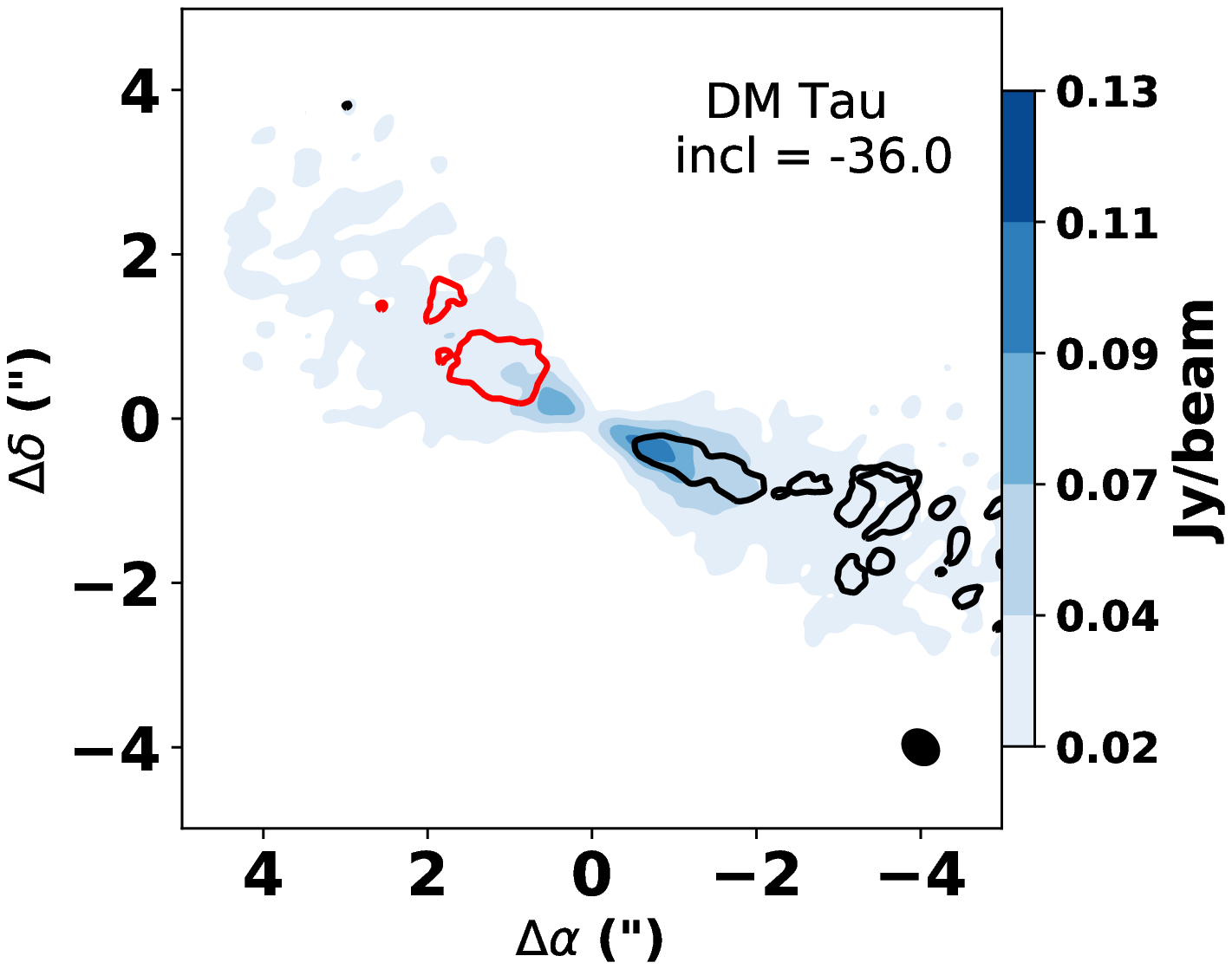}
\includegraphics[scale=.4]{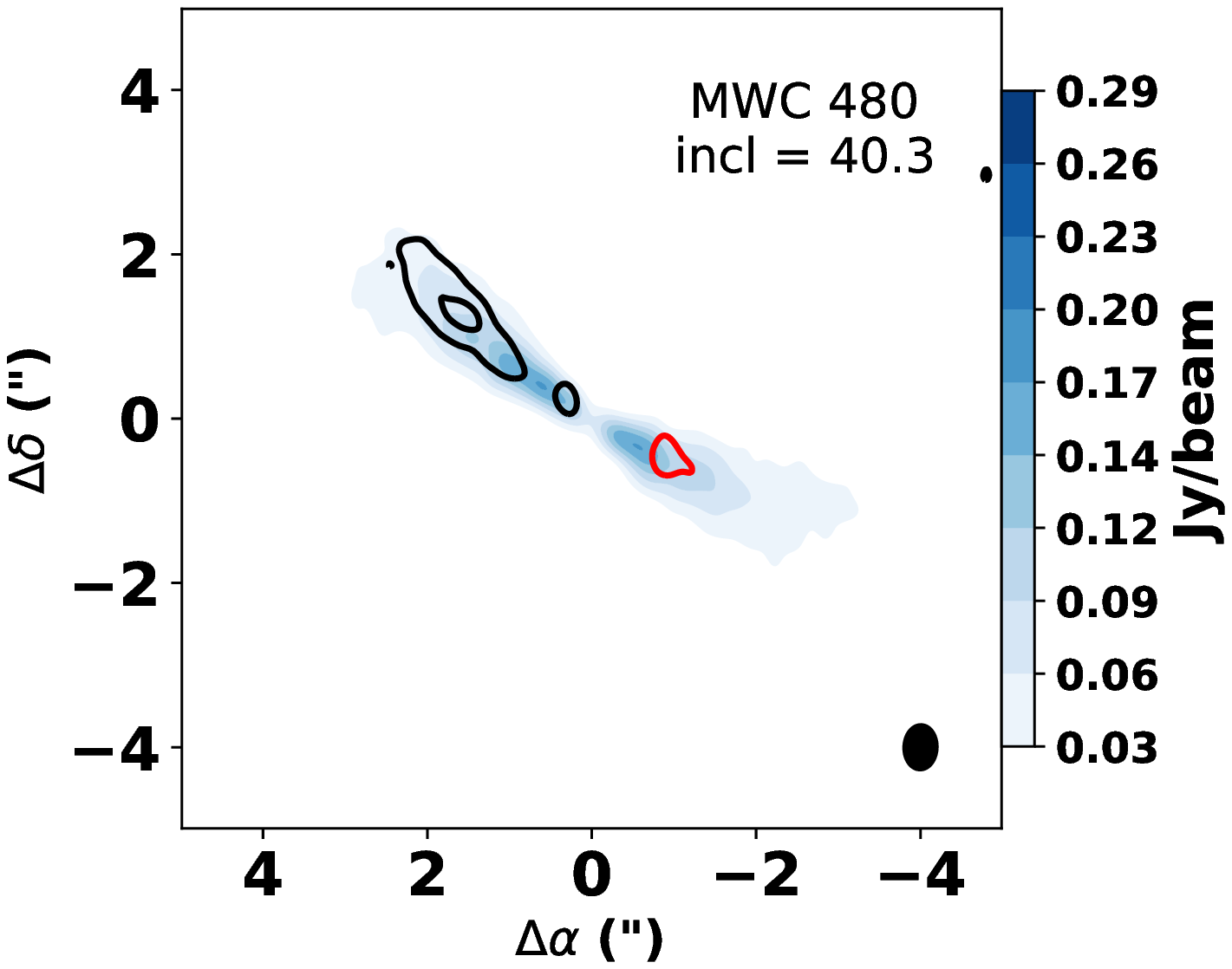}
\includegraphics[scale=.4]{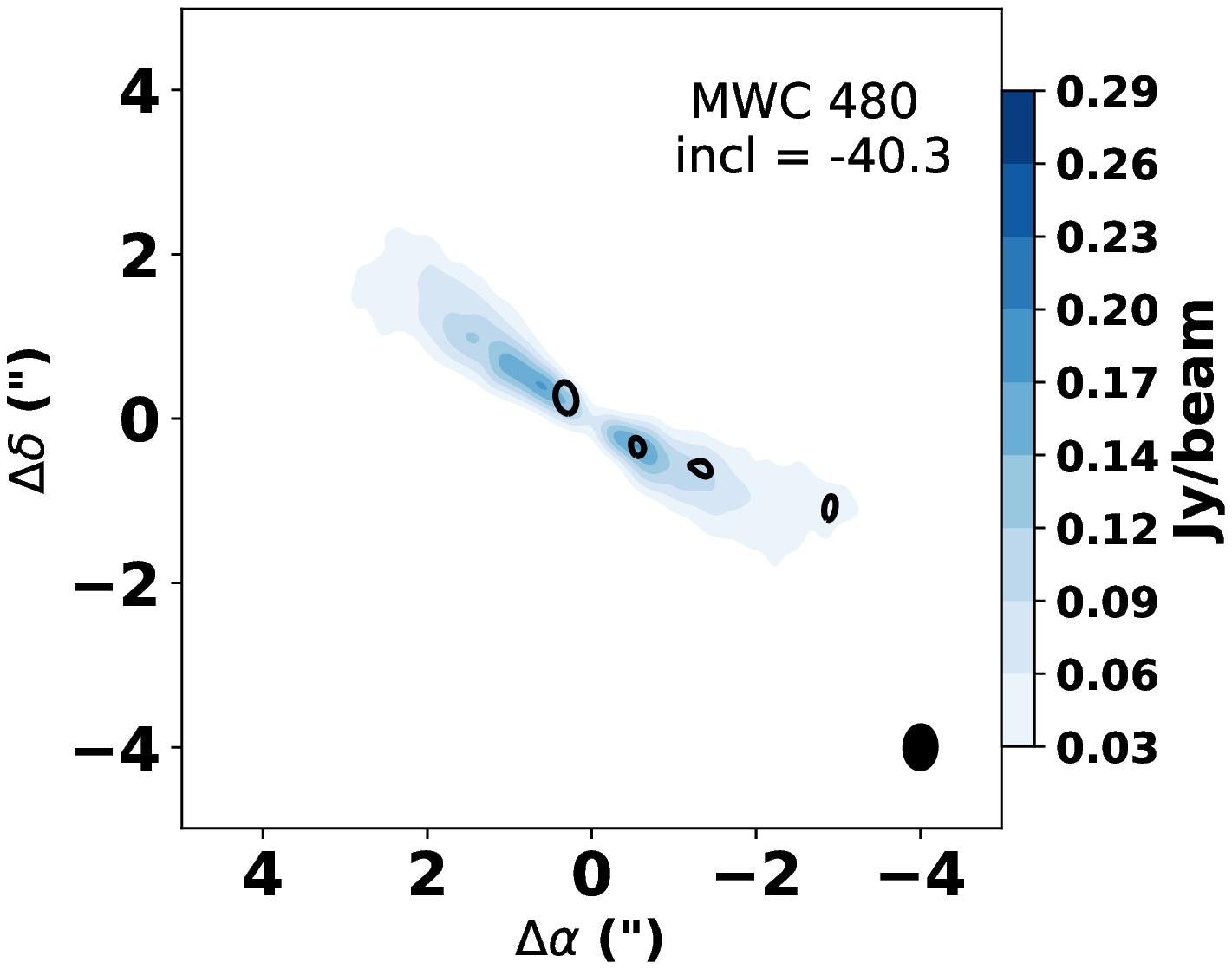}
\includegraphics[scale=.4]{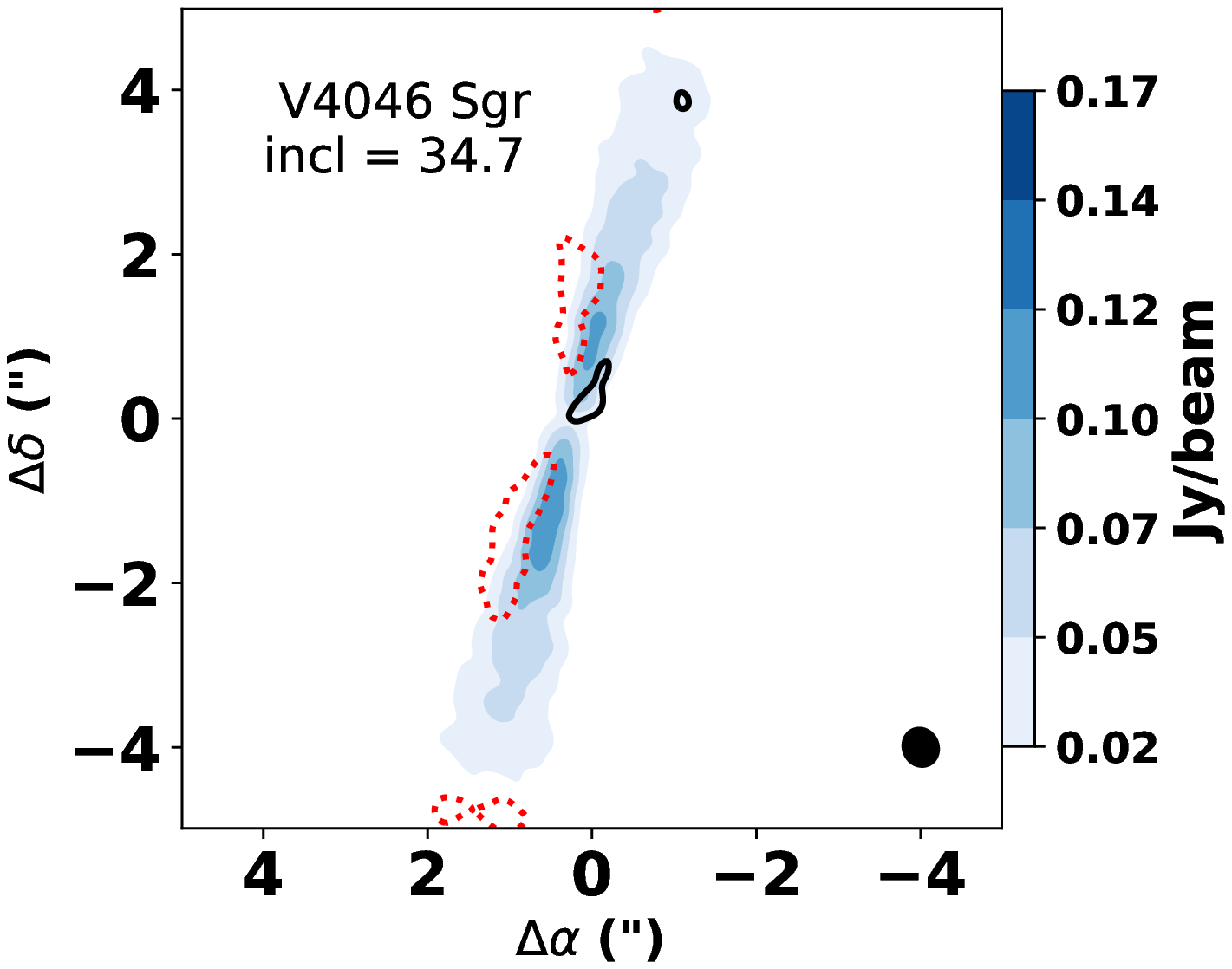}
\includegraphics[scale=.4]{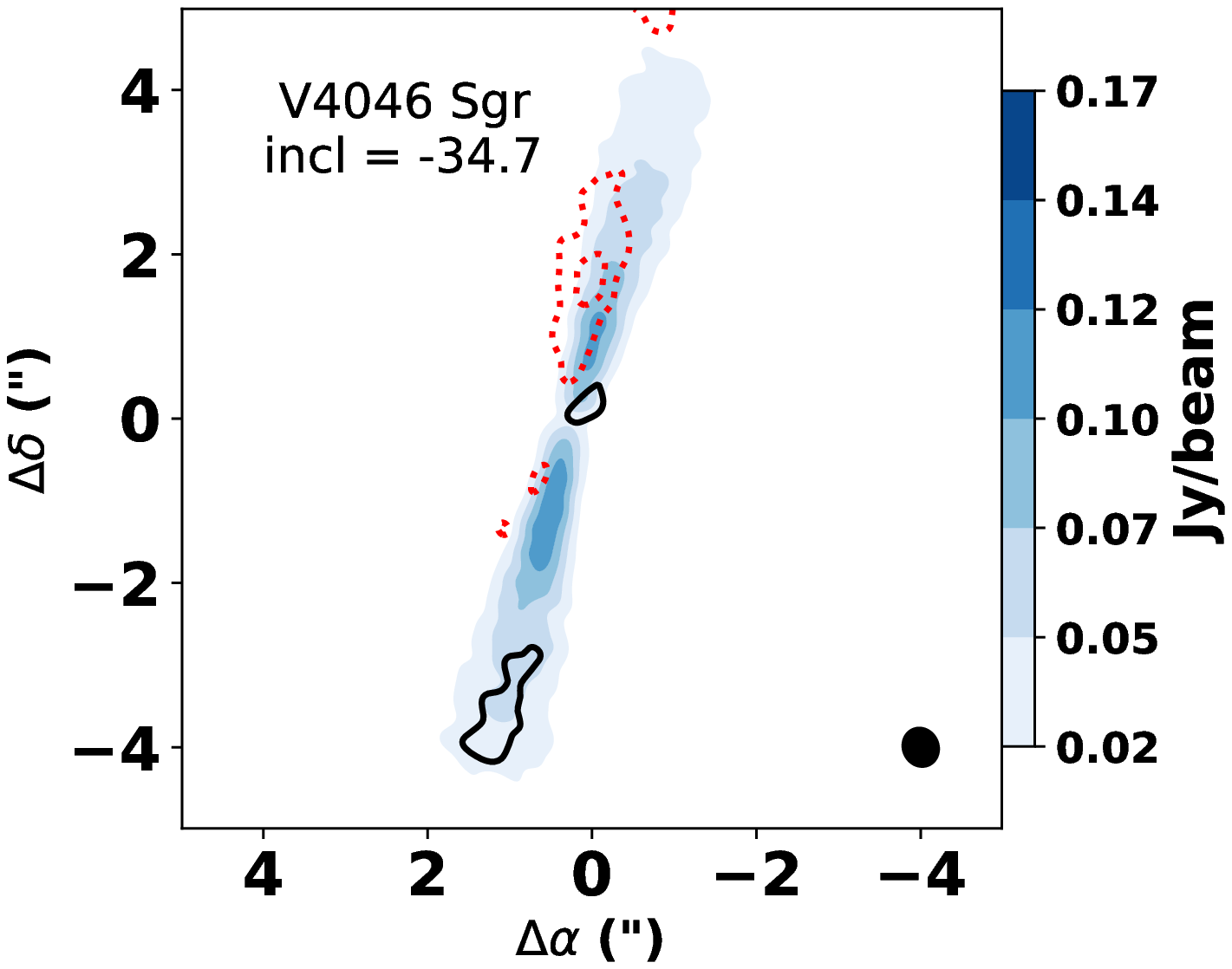}
\caption{The nonzero thickness of the disk introduces a small asymmetry along the minor axis between a model with a positive inclination and a model with negative inclination. Given the high S/N of our data we can distinguish between these scenarios in the central velocity channel maps shown here. The residuals (black contours for data$>$model, red dotted contours for model$>$data) for negative inclinations (left column) show a clear asymmetry along the minor axis for the disks around V4046 Sgr and DM Tau but not for MWC 480. This resolves the degeneracy in the sign of the inclination, and indicates that for the disk around V4046 Sgr the north-west side is closer to the observer, for DM Tau the north-east side of the disk is closer, and for MWC 480 the south-west side of the disk is closer.\label{incl_degeneracy}}
\end{figure*}

Figure~\ref{incl_degeneracy} shows the residuals for the three systems between models with a positive and negative inclination \citep[following the disk orientation convention outlined in ][]{cze19}. The disks around DM Tau and V4046 Sgr exhibit residuals that are asymmetric along the minor axis of the disk when the inclination is negative, while these residuals disappear when the inclination is positive. Conversely, the disk around MWC 480 has asymmetric residuals for a positive inclination, but more symmetric residuals for a negative disk inclinations. This is evidence that the disks around DM Tau and V4046 Sgr have positive inclinations (i.e. the north-east and north-west regions of the disk around DM Tau and V4046 Sgr, respectively, are closer to the observer), while the disk around MWC 480 has a negative inclination (i.e. the south-west region of the disk is closer to the observer than the north-east).

\section{Posterior Distribution Functions\label{pdfs}}
\begin{figure*}
\center
\includegraphics[scale=.45]{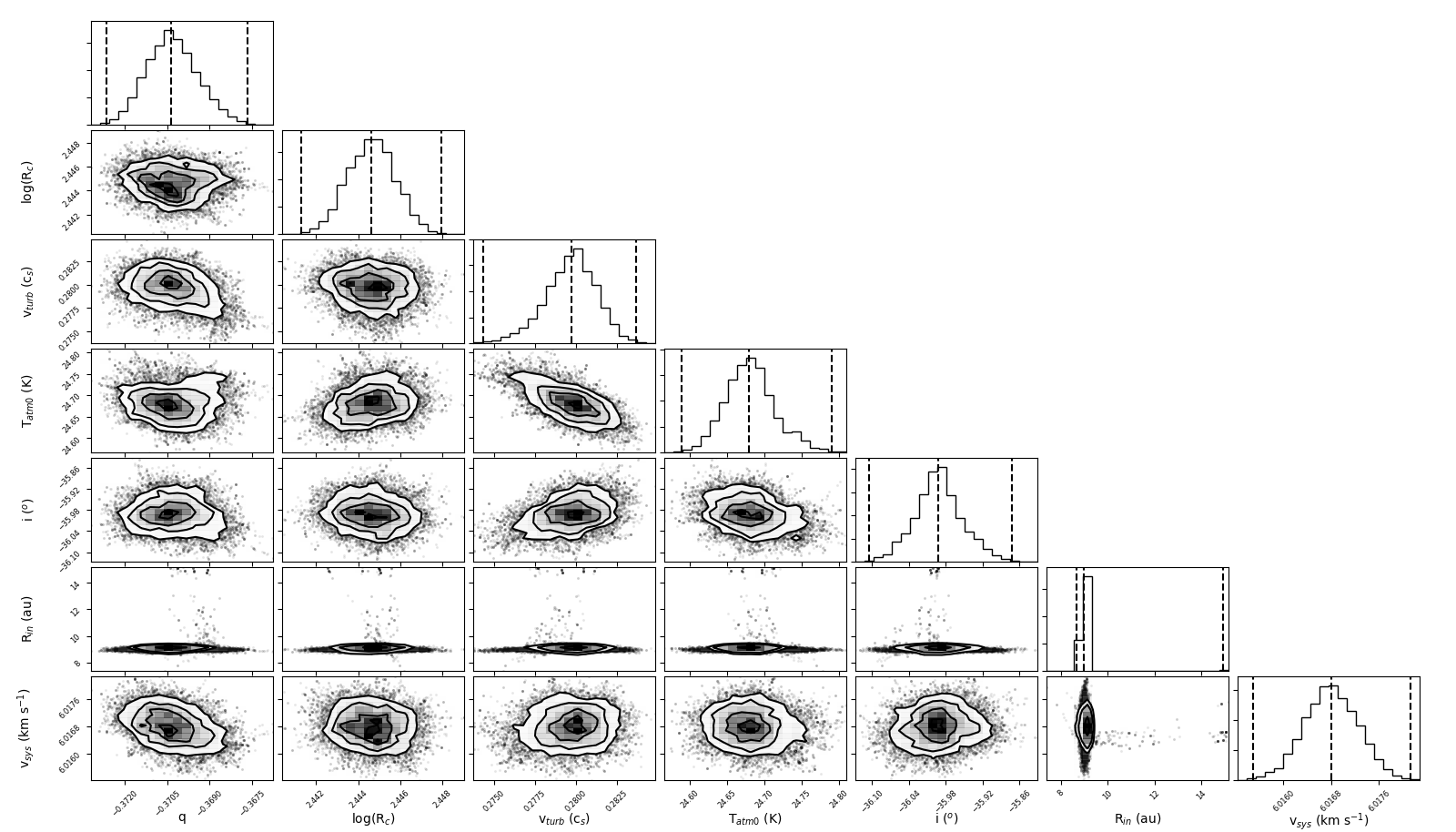}
\caption{Corner plot showing one and two-dimensional posterior distributions for the DM Tau fiducial model. Vertical lines indicate the 1.5, 50, 99.85 percentile locations. \label{dmtau_pdfs}}
\end{figure*}

\begin{figure*}
\center
\includegraphics[scale=.45]{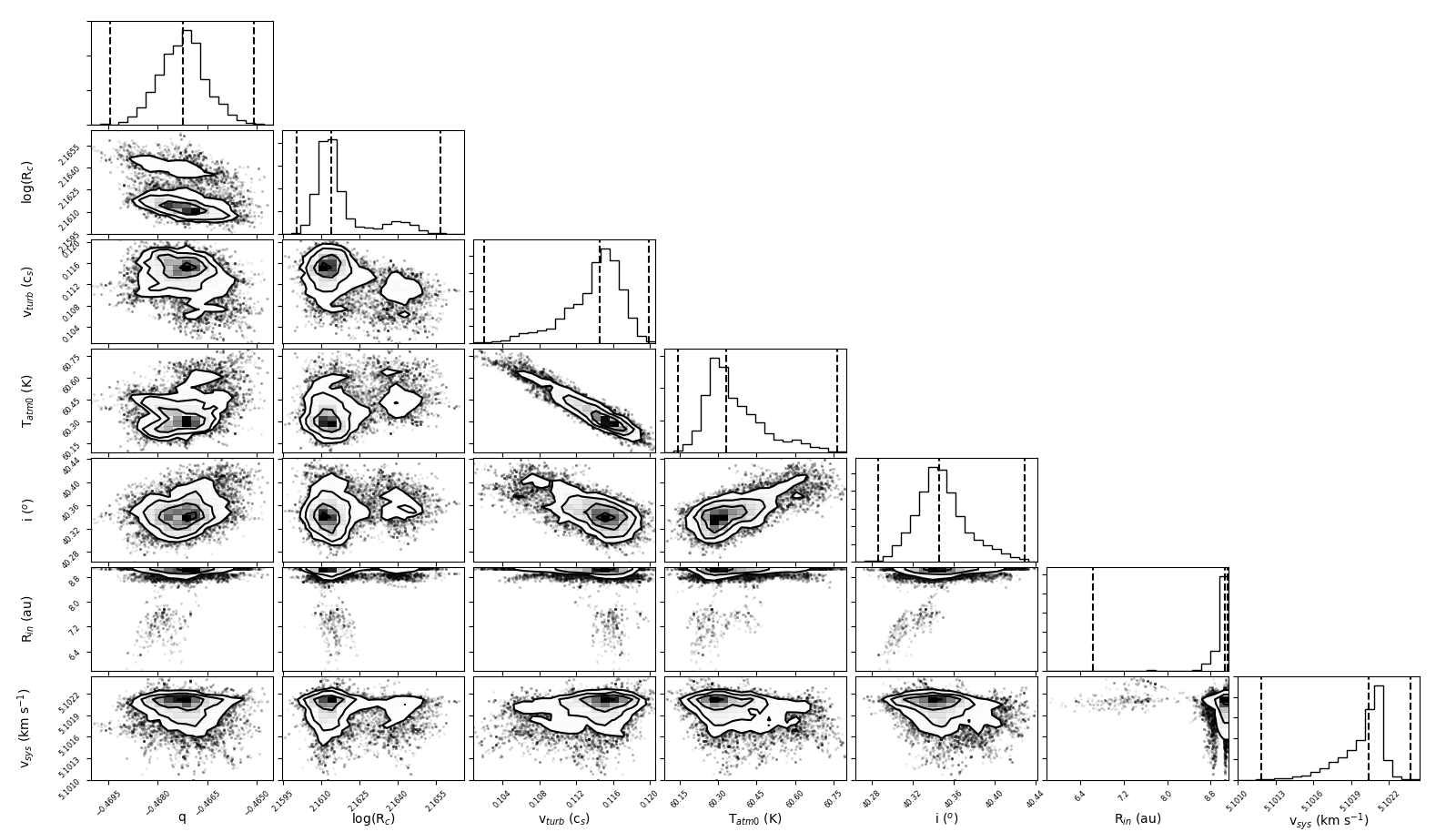}
\caption{Corner plot showing one and two-dimensional posterior distributions for the MWC 480 fiducial model. Vertical lines indicate the 1.5, 50, 99.85 percentile locations. There is a degeneracy between $T_{\rm atm0}$ and $v_{\rm turb}$, although it extends over a small range in parameter space. \label{mwc480_pdfs}}
\end{figure*}

\begin{figure*}
\center
\includegraphics[scale=.45]{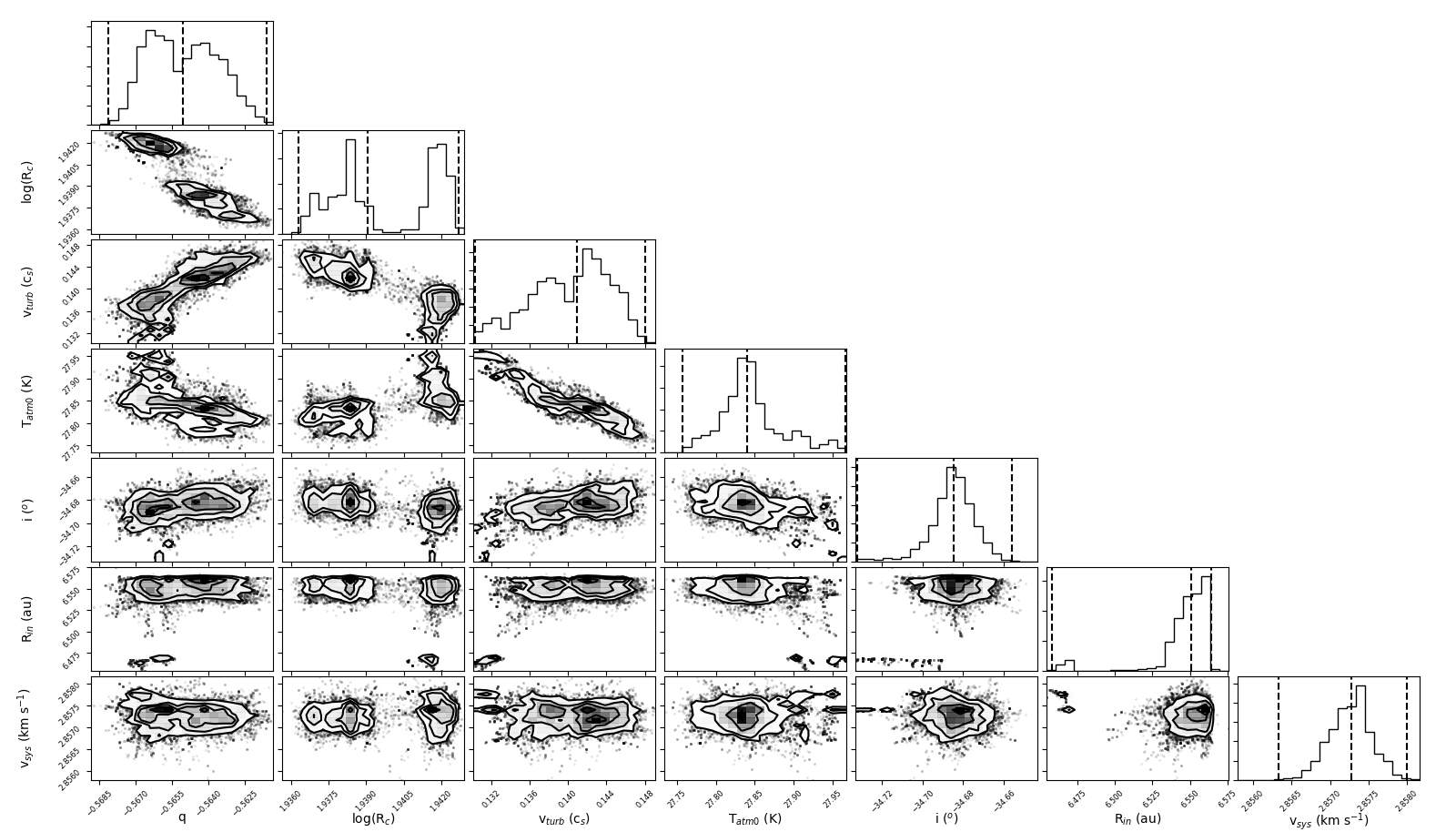}
\caption{Corner plot showing one and two-dimensional posterior distributions for the V4046 Sgr fiducial model. Vertical lines indicate the 1.5, 50, 99.85 percentile locations. There is a degeneracy between $T_{\rm atm0}$ and $v_{\rm turb}$, although it extends over a small range in parameter space. \label{v4046_pdfs}}
\end{figure*}

\begin{figure*}
\center
\includegraphics[scale=.47]{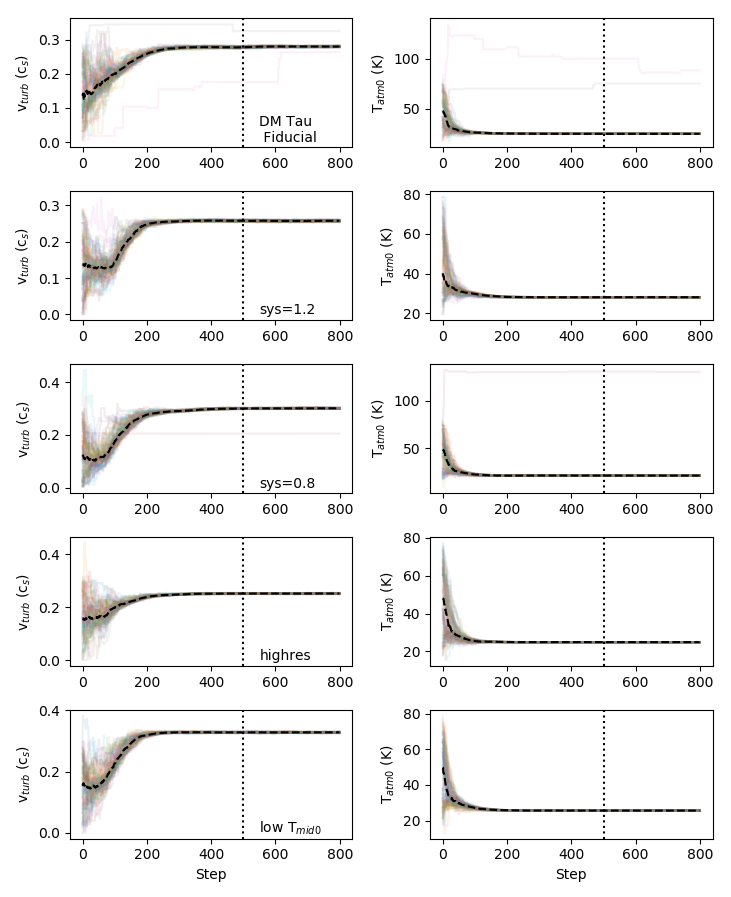}
\includegraphics[scale=.47]{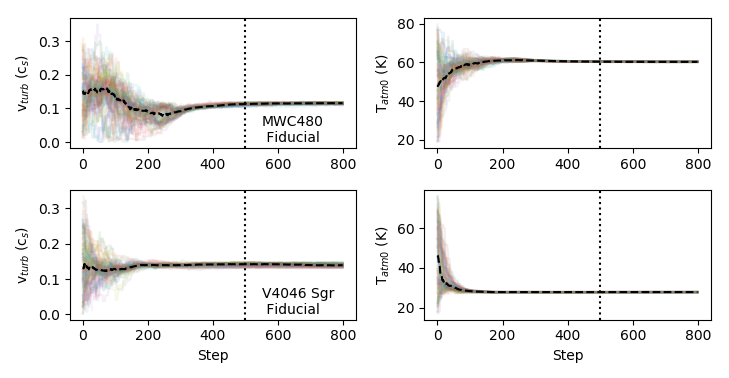}
\caption{(Left): Progress of the walkers for the $v_{\rm turb}$ and $T_{\rm mid0}$ parameters for the DM Tau trials. Individual walkers are indicated by the light colored lines, while the median is shown by the black dashed line. The vertical dotted line divides burn-in from the region used in generating the PDFs. In each trial, the walkers quickly converge toward a preferred solution; walkers that do not converge are excluded when generating the PDFs. Note that the difference in results between the models is much larger than the statistical dispersion within a particular trial, indicating that systematic uncertainties are much larger than statistical uncertainties. (Right): Walker progressions for MWC 480 (top) and V4046 Sgr (bottom). As with DM Tau, the walkers quickly converge to a narrow range in parameter space. \label{dmtau_chains}}
\end{figure*}

Here we show the two-dimensional and one-dimensional posterior distribution functions for the fiducial MCMC trials for DM Tau (Figure~\ref{dmtau_pdfs}), MWC 480 (Figure~\ref{mwc480_pdfs}), and V4046 Sgr (Figure~\ref{v4046_pdfs}).  Figure~\ref{dmtau_chains} shows the chains for DM Tau trials, demonstrating that the walkers quickly converge toward a preferred solution. 

\end{document}